\documentclass[preprint2]{emulateapj}

\usepackage{gensymb}

\usepackage{subfigure}
\newcommand{\comment}[1]{}

\begin{document}

\title{Spectroscopic Observations of Obscured Populations in the Inner Galaxy: 2MASS-GC02, Terzan~4, and the 200~km~s$^{-1}$ stellar peak\footnote{Based on observations taken by APOGEE}}

\author{
Andrea Kunder\altaffilmark{1},
Riley E. Crabb\altaffilmark{1},
Victor P. Debattista\altaffilmark{2},
Andreas J. Koch-Hansen\altaffilmark{3},
Brianna M. Huhmann\altaffilmark{1}
}
\altaffiltext{1}{Saint Martin's University, 5000 Abbey Way SE, Lacey, WA, 98503, USA}
\altaffiltext{2}{Jeremiah Horrocks Institute, University of Central Lancashire, Preston, PR1 2HE, UK}
\altaffiltext{3}{Zentrum f\"ur Astronomie der Universit\"at Heidelberg, Astronomisches Rechen-Institut, M\"onchhofstr. 12-14, 69120 Heidelberg, Germany,}

\begin{abstract}
The interpretation of potentially new and already known stellar structures located at low-latitudes 
is hindered by the presence of dense gas and dust, as observations toward these sight-lines 
are limited.  We have identified APOGEE stars belonging to the low-latitude globular 
clusters 2MASS-GC02 and Terzan~4, presenting the first chemical element abundances of stars 
residing in these poorly studied clusters.  As expected, 
the signature of multiple populations co-existing in these metal-rich clusters is evident. 
We redetermine the radial velocity of 2MASS-GC02 to be $-$87 $\pm$ 7~km~s$^{-1}$, finding 
that this cluster's heliocentric radial velocity is offset by more than 150~km~s$^{-1}$ from the 
literature value.  
We investigate a potentially new low-latitude stellar structure, a kiloparsec-scale nuclear disk (or ring) which 
has been put forward to explain a high-velocity ($V_{GSR}$ $\sim$200~km~s$^{-1}$) peak 
reported in several Galactic bulge fields based on the APOGEE commissioning observations.  
New radial velocities of field stars at ($l$,$b$)=($-$6$^\circ$,0$^\circ$) are presented and combined with the 
APOGEE observations at negative longitudes to carry out this search.
Unfortunately no prominent $-$200~km~s$^{-1}$ peak at negative longitudes along the plane 
of the Milky Way is apparent, as predicted for the signature of a nuclear feature.  
The distances and {\it Gaia} EDR3 proper motions of the high-$V_{GSR}$ stars do not 
support the current models of stars on bar-supporting orbits as an explanation of 
the +200~km~s$^{-1}$ peak. 
\end{abstract}


\section{Introduction} 
\label{sec:intro}
Mapping the structure of the Galaxy is an essential endeavor, as the Milky Way's (MW) assembly 
and star formation history is encoded in the kinematics, metallicities, ages and spatial 
distribution of its stars \citep[e.g.,][]{freeman02, cooper10, minchev16, helmi18}.  The kinematics and composition of 
stars in the MW allow for the interpretation of individual components within the Galaxy, 
and relate these to the fundamental properties of galaxy formation models. 

The Galactic plane is a region that has remained particularly difficult to probe, as 
extreme obscuration by interstellar dust and gas, high source density, 
and confusion with foreground disk populations raise significant limitations in 
any kind of study \citep{gonzalez12, nataf16, alonsogarcia18, nogueraslara19}.  
Consequently, our view of the inner Galaxy and Galactic bulge is 
dominated by observations of stars at Galactic latitudes larger
than $|b|\sim$2$^\circ$. 

Modern large surveys using new instrumentation on large telescopes 
are beginning to penetrate stellar populations that exist also along the plane of the bulge.  
For example, the infrared photometric VISTA Variables in the V\'{i}a L\'{a}ctea (VVV) ESO 
Public Survey \citep{minniti10} and the spectroscopic Apache Point Observatory Galaxy Evolution 
Experiment \citep[APOGEE,][]{eisenstein11} are covering large areas of the plane ($|b|<2^\circ$)
of the bulge.  This has allowed detailed studies of stars in this populous and crowded 
region in a more focused fashion.  
 
However, one well-known gap in the literature concerns the population of globular clusters (GCs) 
in the bulge near the plane of the Galaxy.  
Surveys such as APOGEE have concentrated more on the GCs in the outskirts of the Galaxy 
than the bulge \citep[e.g.,][]{meszaros20}, with the bulge GCs needing to be treated in a more 
careful manner \citep[e.g.,][]{schiavon17,fernandez19,fernandez20a}.  This is unfortunate as the bulge GCs are 
important probes of the formation processes of the central parts of the Galaxy \citep[e.g.,][]{barbuy18}.
There is also a growing debate on the 
number of globular clusters that exist in the inner Galaxy, as discriminating real from 
spurious clusters candidates is troublesome \citep[e.g.,][]{froebrich07, gran19, bdbs1}.  

Another ongoing issue concerns the so-called 200~km~s$^{-1}$ peak, which has been 
reported in the APOGEE fields closest to the plane of the Galaxy.
This cold stellar feature ($\sigma_{V}~\sim$35~km~s$^{-1}$) with a 
galactocentric velocity ($V_{GSR}$) of $\sim$200 km~s$^{-1}$ was first reported in 
several Galactic bulge fields 
based on the APOGEE commissioning observations \citep{nidever12}.  
\citet{nidever12} originally proposed that the 
cold  high-$V_{GSR}$ stars are on bar orbits.  

Both \citet{li14} and \citet{gomez16}
showed that bar-supporting orbits would produce a high-velocity shoulder, but no discrete peak.  
Although indeed, many of the APOGEE fields exhibit shoulders and not peaks, 
3 of the 53 of the APOGEE fields at positive longitudes do show a discrete $V_{GSR}$ peak 
with a clear trough \citep{zhou17}.  For the negative longitudes, \citet{zhou20} find no 
distinct cold high-$V_{GSR}$ peaks in the APOGEE fields, instead finding that complex 
velocity distributions fit these fields best.  However, observations at the negative longitudes 
are not as extensive or complete, hampering a detailed study.

\citet{aumer15} argued that the APOGEE selection function favors young stars, and
younger stars would be trapped by the bar into resonant orbits and could give rise to the 
cool, high-velocity peaks.
However, \citet{zasowski16}, \citet{zhou17} and  \citet{zhou20} showed that stars in the
APOGEE high-$V_{GSR}$ peaks do not exhibit distinct chemical
abundances or ages, indicating that they are not predominantly comprised of younger stars.
Recently, \citet{mcgough20} show that stars in the propellor orbit family, in isolation, have a 
kinematic signature similar to that of the 200~km~s$^{-1}$ peak.

A further explanation of the cold, high-velocity peak is that the stars in the high-$V_{GSR}$ peak are
the signature of a kiloparsec-scale nuclear disk (or ring) that could exist at the center
of the Milky Way \citep{debattista15}.  In this interpretation, the high-velocity peak 
seen clearly in the APOGEE data at ($l$,$b$) = (6$^\circ$,0$^\circ$) is the tangent point 
of a nuclear structure supported by $\rm x_2$ orbits, which are perpendicular to the bar \citep[see $e.g.$, Figure~1,][]{debattista15}.   We stress that the  kiloparsec-scale nuclear disk put forward by this model 
is unrelated to the nuclear stellar disk -- a dense, disk-like complex of stars and molecular clouds 
in the nuclear bulge \citep{launhardt02, nishiyama13, nogueraslara20} with a radius of $\sim$150 pc. 

Here we use the recently released SDSS-IV APOGEE-2 database  \citep{ahumada20} supplemented by 
our own new radial velocities at ($l$,$b$) = ($-$6$^\circ$,0$^\circ$) to search for 
the signature of the high-velocity peak at negative longitudes.  In this way, we can see if there 
is further supporting evidence of a nuclear feature in the Galaxy.
Due to the geometry of the disk, the cold, high-velocity peak should be more pronounced on the 
$l <$ 0$^\circ$ side if this is due to a nuclear disk or ring \citep{debattista18}.  Therefore 
the detection of a high-velocity peak at negative longitudes should be as apparent, if not 
more, using stars at negative longitudes of the Galaxy. 

Unfortunately, very little spectroscopy exists for bulge stars at negative longitudes along the plane 
of the Milky Way.  \citet{trapp18} reported a tentative signature of a ring using SiO masers at 
$l \sim -8.5^\circ$, but their small sample ($\sim$140 masers) and large uncertainty in the velocity 
position prevented a definite conclusion, let alone to determine chemical abundances.   

In \S2, we describe the data used in this analysis of the plane of the Galaxy.  We test the reliability 
of using {\it Gaia} proper motions in our analysis by considering the proper motions of individual stars 
belonging to various globular clusters in \S3.  Statistical tests to quantify the signature of a high-velocity 
peak at negative longitudes are carried in \S4, and conclusions are given in \S5.

\section{Data}
\subsection{APOGEE}
The Apache Point Observatory Galactic Evolution Experiment (APOGEE) is a portion of the 
Sloan Digital Sky Survey (SDSS) III, and has provided a range of data for $\sim$4700 K/M-giant stars 
in the Milky Way bulge \citep{majewski16}.  As APOGEE uses the 2.5m Sloan telescope in the 
Northern hemisphere, most of the bulge observed is at positive longitudes.  APOGEE is 
able to collect near infrared spectra for $\sim$300 targets simultaneously.  Each of their fields are 
observed for roughly one hour which produces high signal-to-noise (SNR) 
spectra ($\rm SNR >100$).

Recently, the successor of APOGEE, SDSS-IV APOGEE-2, has released new observations of 
bulge stars \citep{ahumada20}.  These observations come from the 2.5m du Pont Telescope 
at Las Campanas Observatory, using a twin spectrograph to APOGEE \citep{gunn06}.  
The APOGEE observations give 
estimates of stellar parameters good to $\sim$150 K for $\rm T_{eff}$, 0.2 dex for log $g$ 
and 0.1 dex for $\rm [Fe/H]$ and up to 15 chemical elements have a typical precision 
of 0.1 dex \citep{garciaperez16}.  The APOGEE data releases comprise the largest near 
infrared high-resolution sampling of giant stars of the central region of the Galaxy.  

Distances to the APOGEE stars are provided using the the Bayesian isochrone-fitting code 
StarHorse \citep{queiroz20}.  This code uses spectroscopic data as well as photometric magnitudes and 
{\it Gaia} parallaxes to derive distance estimates with typical uncertainties of $\sim$6\% for APOGEE giants.

To select APOGEE stars that are members of the bulge as opposed to foreground stars, we select those 
stars with surface gravities of $\rm log~g~>~3.8$; this cut is to avoid dwarfs in the foreground.  
%
We also require $(J - K_s)_0 >$ 0.5 \citep[e.g.,][]{nidever12}, where we use the {\tt AK\_TARG} 
value given in 
APOGEE\footnote{APOGEE 
reddening values are calculated using the RJCE method as described in \citet{majewski11}.} 
and the \citet{nishiyama09} extinction law of $A_{Ks}$ = 0.528~$E(J-K_S)$.

\begin{figure}
\centering
\mbox{\subfigure{\includegraphics[height=9.0cm, angle=270]{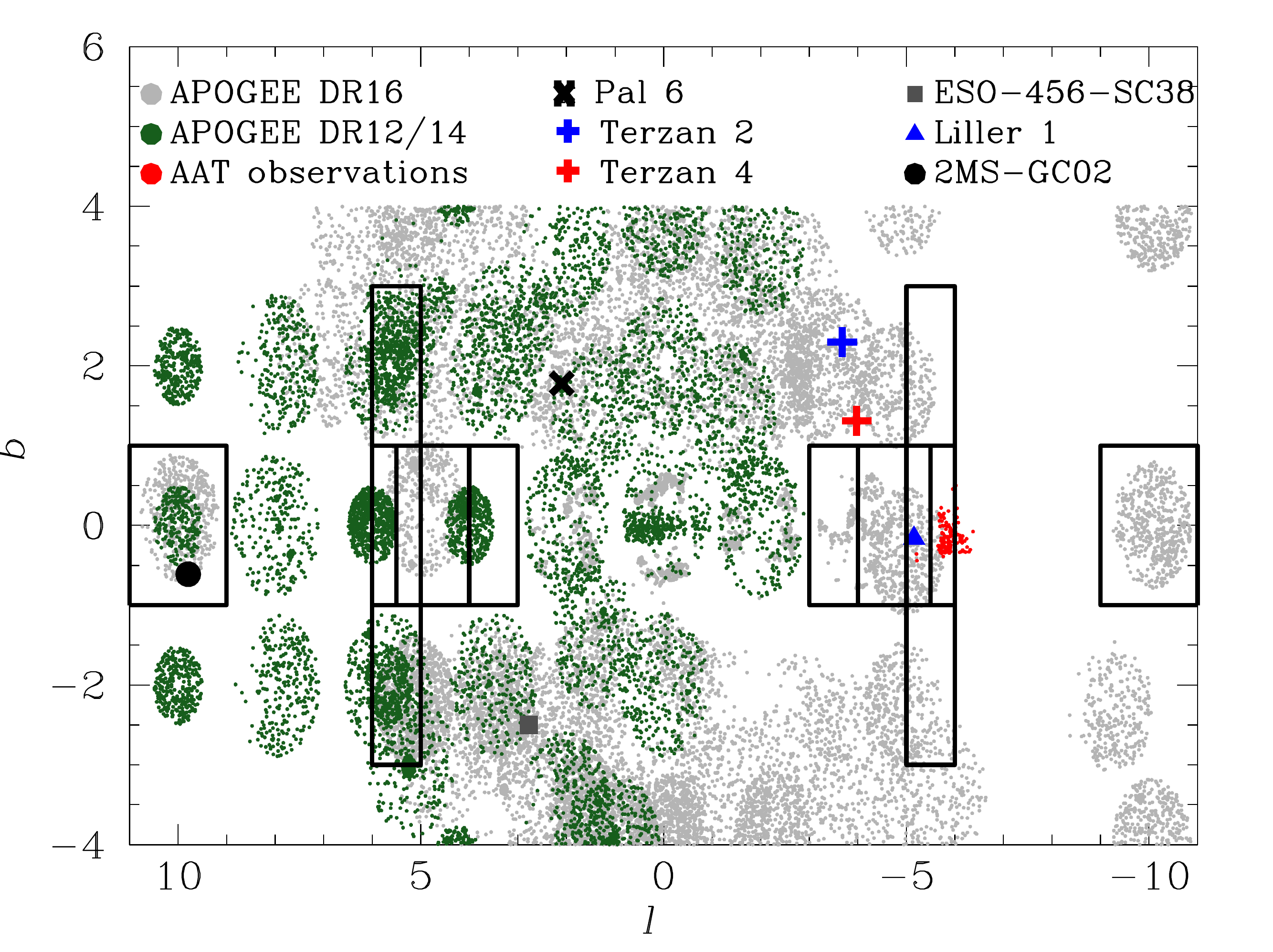}}}
\caption{The spatial location of the APOGEE stars included in this study are shown.  
The grid indicates the spatial division of stars we used to seek for high-$V_{GSR}$ 
($|V_{GSR}| \sim$200~km~s$^{-1}$) peaks.  The increase in spectroscopic spatial 
coverage due to APOGEE DR16 is apparent; in 
particular, the bulge at negative longitudes can now been explored.  Our new AAT 
observations are shown at ($l$,$b$)=($-$6,0), as well as the globular cluster stars 
included in our analysis.}
\label{footprint}
\end{figure}

The results presented in this paper come from the APOGEE giant stars contained within the 
grid shown in Figure~\ref{footprint}.  These observations cover primarily the plane of the 
inner bulge ($|b| <$ 1$^\circ$).  The grid indicates our divisions used in \S4 
(in particular the $V_{GSR}$ histograms shown in Figure~\ref{apogee_flds}) 
to search for potential high-$V_{GSR}$ peaks. 

\subsection{Anglo-Australian Telescope Observations}
One bulge field at ($l$,$b$)=($-$6$^\circ$,0$^\circ$) was observed on June 21, 2020 
(NOAO PropID: 2020A-0368; PI: A. Kunder)  using the AAOmega multifiber spectrograph 
on the Anglo-Australian Telescope (AAT).  As this spectrograph is a dual-beam system, 
we used the 580V grating for the Blue arm and the 1700D grating for the Red arm.  
Due to less than optimal weather conditions, our exposure time was 
2x30 min, after which the clouds rolled in.  Only the spectra from 
the Red arm were used, as this is where the prominent Calcium Triplet lines could still 
be distinguished.  These spectra covered the wavelength regime of about 8350\AA~to 8850\AA~at a 
resolution of R$\sim$10,000.  

The automated pipeline supplied by AAOmega, 2dfdr \citep{aao2dfdr} was used for the data 
reduction.  Fortunately 
we were observing during dark time, so despite some cirrus, our sky subtraction was adequate, 
and our spectra ranged in SNR from $\sim$1 to 10 per pixel.

\begin{figure}
\centering
\mbox{\subfigure{\includegraphics[height=8.0cm]{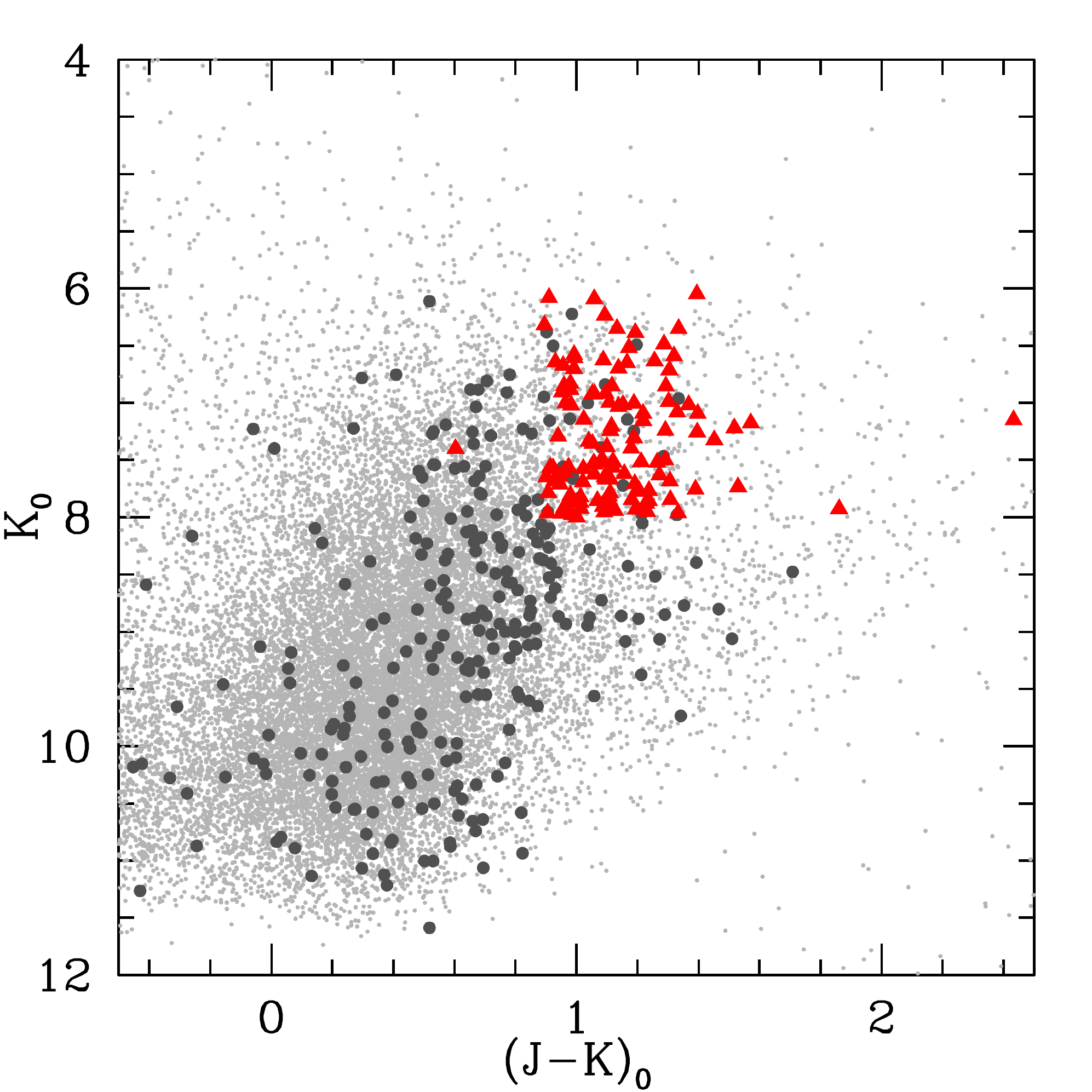}}} 
\caption{The dereddened 2MASS color-magnitude diagram showing the 130 giants (triangles) for which radial 
velocities have been determined from AAOmega@AAT.  The APOGEE stars observed 
at ($l$,$b$)=($-$5.5$^\circ$, 0$^\circ$) are also shown (grey bold points), as are the underlying 
2MASS stellar distribution in this field (small grey points).  
All stars have been dereddened using the extinctions from \citet{gonzalez12} and the \citet{nishiyama09} 
extinction law.  The uncertainties in the reddening and extinction law dominate the errors in 
this plot and range from $\sim$ 0.1 to 0.7~mag in both  $\Delta E(J-K)$ and $\Delta A_K$.
} 
\label{cmd}
\end{figure}

The stellar targets were selected from the 2MASS catalog to be 
located in the ($l$,$b$)=($-$6$^\circ$,0$^\circ$) 
region of the sky that had reliable photometric quality indicators 
{$e.g.,$ \tt ph\_qual=AAA}.  The BEAM extinction calculator \citep{gonzalez12} was used to de-redden all stars.
Figure~\ref{cmd} shows a clearly defined bulge red giant branch 
in this field and supports our selection method to target bulge stars.  The blue cutoff rejects 
many objects that are closer than the bulge, which have lower reddening and are brighter 
than the red giant branch.  A number of stars reach the red edge of the RGB, which is where the most 
metal-rich red giants will fall.

The APOGEE observations in their field 355+00 are also shown.  Some APOGEE stars fall in a region of 
the CMD where contamination from the disk and foreground is significant, as also evidenced 
by their distances which place them primarily either in the foreground or the near-side of the 
bulge \citep[e.g.,][]{queiroz20}.  However, in general, the APOGEE stars also populate the bulge 
red giant branch uniformly.

The AAT spectra were cross-correlated using the IRAF cross-correlation routine, {\tt xcsoa}, using 
the calibration stars taken with the identical set-up.  Our calibration stars include the following 
APOGEE stars:
2M18264551-1747096 (HRV=$-$69.2 km~s$^{-1}$), 
2M18255968-174935  (HRV=49.6 km~s$^{-1}$), 
2M18205442-1751063 (HRV=97.4 km~s$^{-1}$), 
2M18235228-1653096 (HRV=$-$52.7 km~s$^{-1}$), 
2M18233429-1841586 (HRV=165.8 km~s$^{-1}$), 
which were observed in a previous observing run \citep[see][]{kunder20}, as well 
as the APOGEE stars which are located in the same field as our newly observed stars: 
2M17342312-3335162 (HRV=$-$2.2 km~s$^{-1}$) and 
2M17340454-3332161 (HRV=$-$67.3 km~s$^{-1}$).  
In general, the consistency of our
velocity results are 8~km~s$^{-1}$, which is in agreement with the errors reported 
by {\tt xcsao} for each individual measurement.


\section{Globular Clusters}
With the {\it Gaia} satellite, new astrometric data have become available for approximately a billion stars 
in the Milky Way galaxy \citep{gaia18, gaiaedr3, lindegren20}.  Combining radial velocities with proper motions 
from {\it Gaia} allows for all three velocity components of stellar motion to be determined.  
The possibility of using the {\it Gaia} proper motions, therefore, 
completes the measurements of the space motions of a star.  

{\it Gaia} is ongoing with an anticipated five-year mission lifetime, and the most recent data release 
\citep[EDR3,][]{gaiaedr3} is based on 34 months of observations.  Especially in crowded regions of the sky, 
{\it Gaia} faces challenges with astrometric solutions \citep[e.g.,][]{sanders19}; this will undoubtedly 
improve by end-of-mission.  Here we corroborate the accuracy of {\it Gaia} proper motions in 
the plane of the Galaxy near the bulge by checking the consistency of the proper motions of stars 
in GCs.  Six globular clusters in the APOGEE footprint located in or close to the plane were used  
for this purpose, five explored here and one more taken from \citet{kunder20b}.

Table~\ref{GCTab} lists the individual globular clusters used to verify the reliability of the 
{\it Gaia} EDR3 proper motions.  
Column 1 is the name of the GC, column 2 lists the average proper 
motion for the stars selected as cluster candidates, column 3 lists heliocentric distances as 
reported in the 2010 edition of \citet{harris96}, column 4 lists the radial velocity of the cluster, 
and columns 5 and 6 are the published proper motions for the GCs as reported 
by \citet{rossi15} and \citet{vasiliev21}, respectively.  

Previous papers to isolate globular cluster stars use proper motion as a criterion for 
membership \citep[e.g.,][]{schiavon17, baumgardt19, vasiliev19, horta20}, so we searched for 
cluster stars in Palomar~6, Terzan~2, Terzan~4, Liller~1 and 2MASS-GC02 
using primarily APOGEE radial velocities and the radial distance of a star 
from the center of the cluster.  Also, because globular cluster stars are in general more metal-poor 
than the field population \citep[e.g.,][]{harris96, schiavon17, horta20, meszaros20}, 
the metallicities of the stars were  used to select candidate cluster members.  The published 
metallicity and radial velocity values for the clusters were used as soft boundaries 
to guide selection.  

Twenty-three cluster stars across the 5 GCs were identified based on their radial velocity, 
distance from the center of cluster and when available, metallicity.
Table~\ref{chartable} lists the individual stars in each targeted globular cluster, 
as well as the radial velocity and the elemental 
abundances $\rm [Fe/H], [C/Fe], [N/Fe], [Na/Fe], [Mg/Fe]~and~[Al/Fe]$ of the star 
as reported by APOGEE DR16.  The distances in column 4 are from StarHorse, 
taken directly from \citet{queiroz20}.
While APOGEE stars in the cluster Palomar 6 were 
studied previously in \citet{schiavon17}
and \citet{meszaros20}, we still carried out a selection of stars in this cluster.  This was 
to avoid any kind of potential proper motion biases that would artificially lower the scatter 
in the average proper motions of these stars.  We present the first APOGEE stars in the 
globular clusters 2MASS-GC02, Terzan~4 and Liller~1.
\subsection{2MASS-GC02}
\begin{figure}
\centering
\mbox{\subfigure{\includegraphics[width=4.2cm]{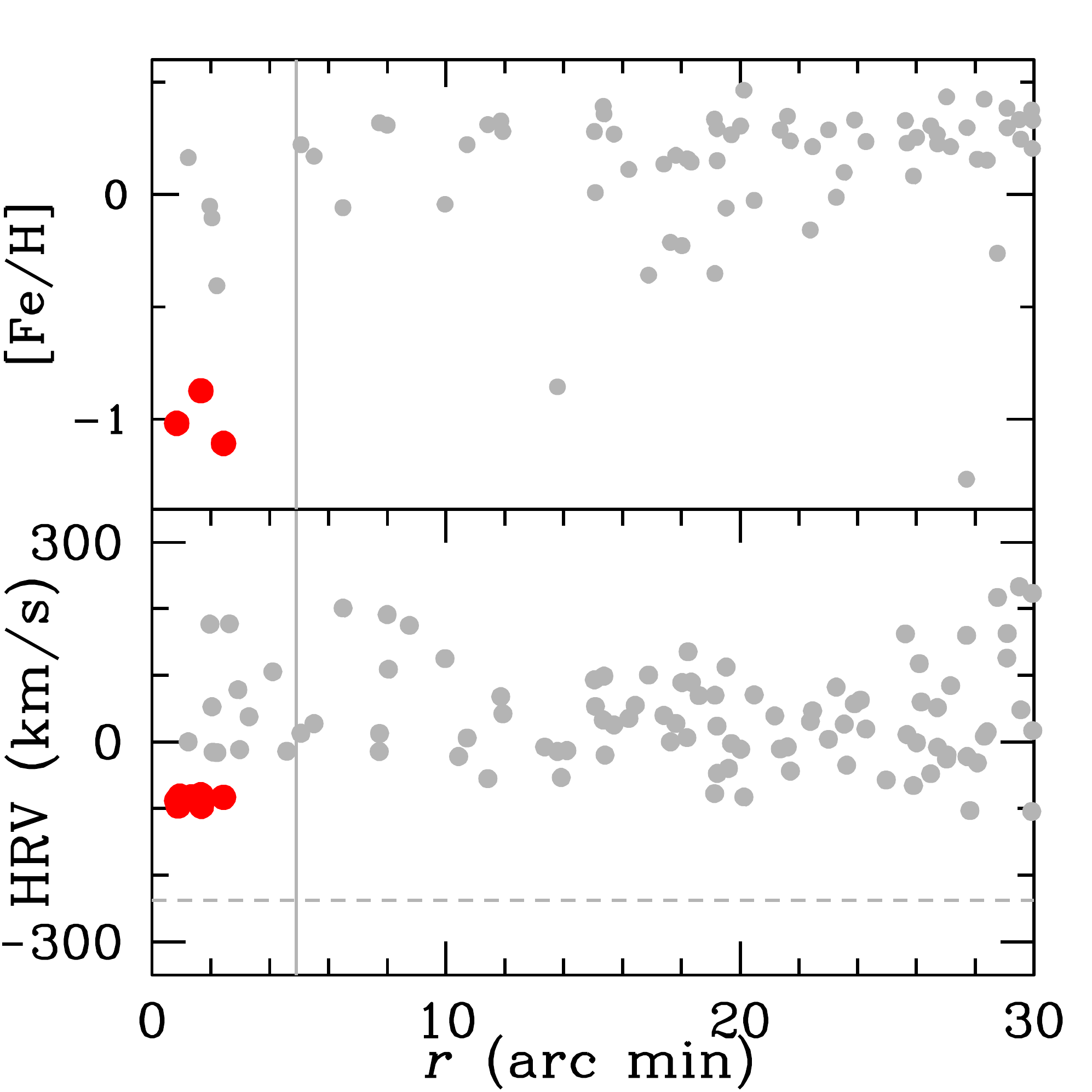}}}
{\subfigure{\includegraphics[height=4.2cm]{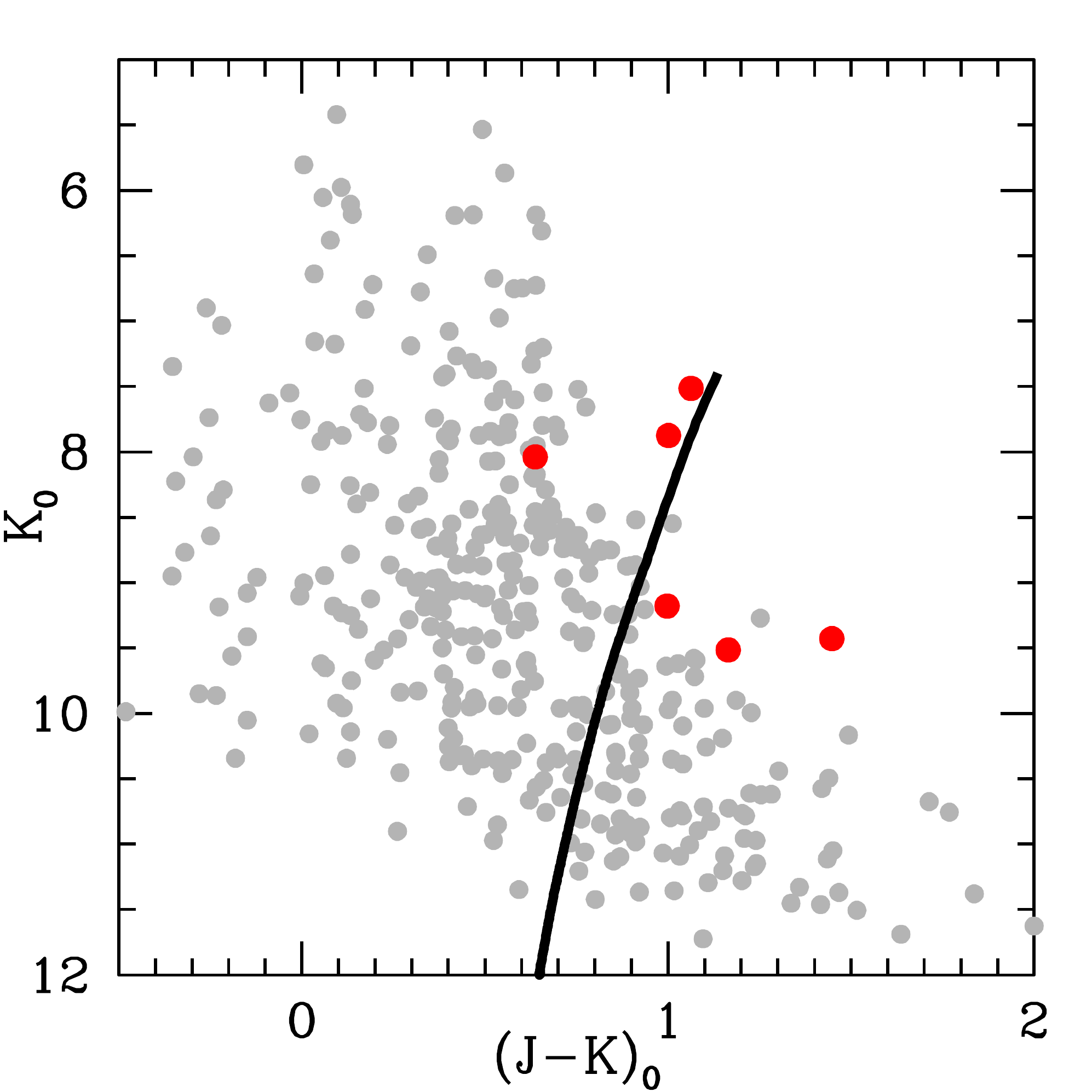}}}
\caption{The heliocentric velocity (bottom left) and $\rm [Fe/H]$ metallicity (top left)
of APOGEE stars within 30 arcminutes of 
2MASS-GC02.  The tidal radius of $r_t$=4.9' is shown, as is the radial 
velocity of the cluster as derived from \citet{borissova07}.  
The color-magnitude diagram of the same stars is shown in the right panel.  
The clump of seven stars with radial velocities of $-$87~km~s$^{-1}$
at a distance of 1' from the center of the cluster 
are consistent with being cluster stars.  The solid black line shows the BaSTI 
isochrone that best matches the cluster's observed parameters (see text).  The 
APOGEE uncertainties are in $\rm [Fe/H]$ are 0.1-0.2~dex, the APOGEE uncertainties 
in radial velocity are $<$0.5~km~s$^{-1}$, and the uncertainties in the dereddened CMD 
range from $\sim$0.1 to 0.7~mag in both $\Delta E(J-K)$ and $\Delta A_K$. 
}
\label{example2MSGC02}
\end{figure}

Figure~\ref{example2MSGC02} shows an example of the selection of APOGEE stars 
in an understudied GC, 2MASS-GC02 \citep{hurt00}.  An over-density of stars close to 
the center of the cluster with radial velocities of $\sim-$87~km~s$^{-1}$ and metallicities 
of $\rm [Fe/H]\sim-$1.0 is apparent.  These are likely not field stars, but instead almost 
certainly belong to 2MASS-GC02.  Curiously \citet{borissova07} present 15 stars with 
SNR $>$ 30 that lie within 1.3 arcmin of 2MASS-GC02 and that have colors and magnitudes 
consistent with the giant branch of 2MASS-GC02.  From 12 of these stars, they find an 
average radial velocity of $\sim -$238$\pm36~$km~s$^{-1}$.  However, 
Figure~\ref{example2MSGC02} shows no stars with such large radial velocities.  

The right panel of Figure~\ref{example2MSGC02} shows that our seven candidate 
2MASS-GC02 stars also have colors and magnitudes consistent with being members 
of the cluster.  The reddening of this cluster is extreme, $A_V~>$15~mag or 
$E(J-K_s)=$3.1$\pm$0.5~mag \citep{ivanov00, hurt00}, and 
it has been noted that the extinction towards this cluster is highly variable (up to 50\% change
in extinction over small regions), and with significant deviations from the standard extinction 
law \citep{alonsogarcia15}.  \citet{penaloza15} have found the stars in this cluster are 
$\alpha$-enhanced, as is typical of old 
globular clusters in the bulge \citep[e.g.,][]{dias16,meszaros20, horta20}. 
The BaSTI \citep{pietrinferni04, pietrinferni06} $\alpha$-enhanced 
stellar evolution models\footnote{http://basti-iac.oa-abruzzo.inaf.it} were adopted to 
indicate the approximate location of the red giant branch of 2MASS-GC02.  
We used 
the publicly available BaSTI isochrone that best matches the cluster's observed parameters, 
one with an age of 12.5 Gyr, a metallicity of $\rm [Fe/H]$ = $-$1.01~dex, and a distance 
modulus of $(m-M)_0$ = 13.48~mag \citep{borissova07}.  

It therefore seems likely that the seven stars we identified as being cluster members 
of 2MASS-GC02 are bona fide cluster members, and redetermine the 
radial velocity of 2MASS-GC02 to be $-$87 $\pm$ 7~km~s$^{-1}$.  
Unfortunately, only one of the seven stars has a {\it Gaia} proper motion, and so our sample 
to check for proper motion consistency is not sufficient.

It is important to stress that our selection differs from that 
of \citet{baumgardt19}, \citet{vasiliev19} and \citet{vasiliev21}, 
who used the distribution of sources in proper motion space as a parameter to select cluster 
members.  These authors then determined the mean proper motions of almost the entire 
known population of Milky Way GCs.  
Because proper motion is not used as a selection criterion for isolating cluster stars here, we 
can then find the scatter in the proper motions of our selected GC stars.  We use this as an 
indicator of the reliability of {\it Gaia} proper motions in a typical sample of APOGEE stars 
residing in the most crowded regions of the inner Galaxy.  

\begin{figure}
\centering
\mbox{\subfigure{\includegraphics[height=8.0cm]{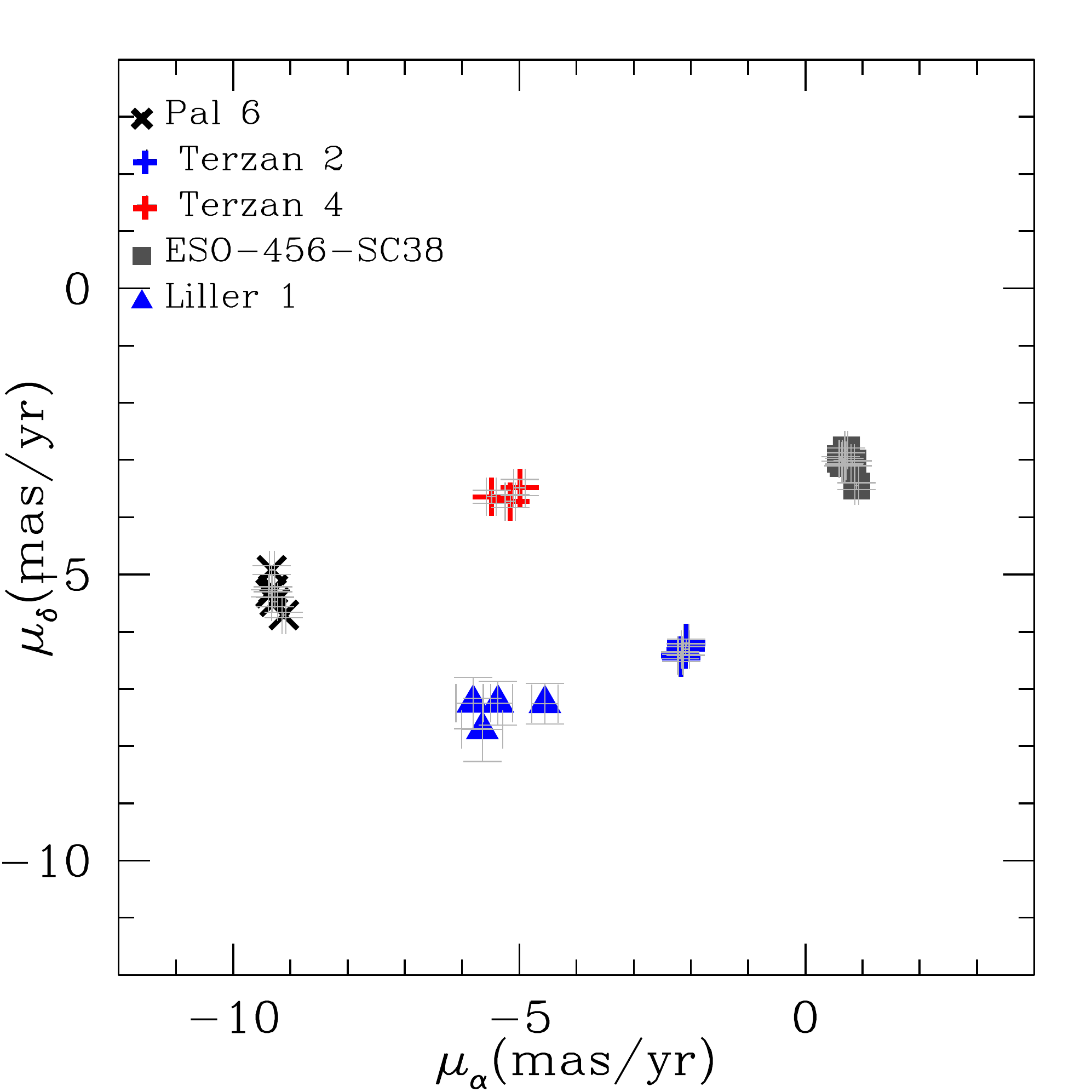}}} %
\caption{The proper motion distribution for globular cluster stars in APOGEE selected 
using radial velocity, metallicity, and distance from cluster center.  The formal errors 
on the proper motions as reported by {\it Gaia} are indicated by thin grey lines and are 
typically smaller than the size of the symbols.  The {\it Gaia} EDR3 proper 
motions are stable also in crowded bulge regions, $e.g.,$ within 2.5 degrees from the plane of the Galaxy.} 
\label{proper_motion_scatter}
\end{figure}
Figure~\ref{proper_motion_scatter} shows the individual proper motions of the cluster stars 
identified here.  
In general, the average values of cluster stars we selected (without using proper motion 
criteria) have a proper motion scatter 
of $\sim$0.15~mas~yr$^{-1}$ or less.  This value is in agreement with the average uncertainty in proper motion 
found in \citet{vasiliev21}, also using globular cluster stars.  Although the absolute proper motion values 
for the clusters are consistent with the values presented 
by \citet{baumgardt19} and \citet{vasiliev19}, they are not consistent with the absolute proper 
motion values presented in \citet{rossi15}, which are computed using field stars as an absolute 
reference frame.  The \citet{rossi15} proper motion values 
dramatically differ in zero-point by 3~mas~yr$^{-1}$ to 11~mas~yr$^{-1}$, and no correlation between 
the {\it Gaia} EDR3 proper motions and those presented by \citet{rossi15} is apparent.  
Although systematic errors 
in {\it Gaia} proper motions have been reported \citep[e.g.,][]{vasiliev_systematics19, lindegren18, gaia_helmi18}, offsets of 
$>$1 mas~yr$^{-1}$ are not thought to be likely.

\subsection{Chemical Element Abundances}
Observational evidence demonstrates that most Galactic GCs host (at least) two main groups of stars with 
different chemical composition.  One population of stars, referred to as the first generation (1G), 
have a similar chemical composition as halo field stars with similar metallicity, while the 
other population of stars (2G) have enhanced helium, nitrogen and sodium abundances, but are 
depleted in carbon and oxygen \citep[e.g.,][]{kraft94, carretta09, marino19}.  
The origin of these multiple populations is still an open issue \citep[e.g.,][]{gratton12, renzini15, bastian18}.

GCs in the inner Galaxy represent a moderately metal-rich population and are thought to 
be some of the oldest GCs in the Galaxy \citep[e.g.,][]{vandenberg13, massari19}.  However, 
bulge GCs are less studied than those in the halo, as high and often variable extinction as well as 
crowding and the difficulty in separating true bulge stars from intervening thin and thick disk stars, 
makes them more difficult to analyze in detail \citep[e.g.,][]{bica16}.  Whereas almost all GCs 
studied have shown signs of multiple-populations with certain dependencies on mass (and age), 
it is less clear how $\rm [Fe/H]$ metallicity affects the multiple-population 
phenomenon \citep{kayser08, carretta09, hanke17, bastian18}.  For example, old, metal-rich 
clusters have been found that do not exhibit multiple 
populations \citep[e.g.,][]{salinas15} and these clusters have been used to suggest 
improvement in the models of GC formation \citep{tang17}.

This paper is the first to present the APOGEE stars and their chemical element abundances in 
the clusters 2MASS-GC02 and Terzan 4.  As such, a search for multiple populations in these 
clusters is carried out.  Due to the small numbers of stars per cluster with elemental abundances, 
we are hesitant to carry out a deeper analysis, $e.g.,$ identifying gaps in the elemental 
distribution or identifying clusters that are chemically analogous.

\begin{figure}
\centering
\mbox{\subfigure{\includegraphics[height=4.25cm]{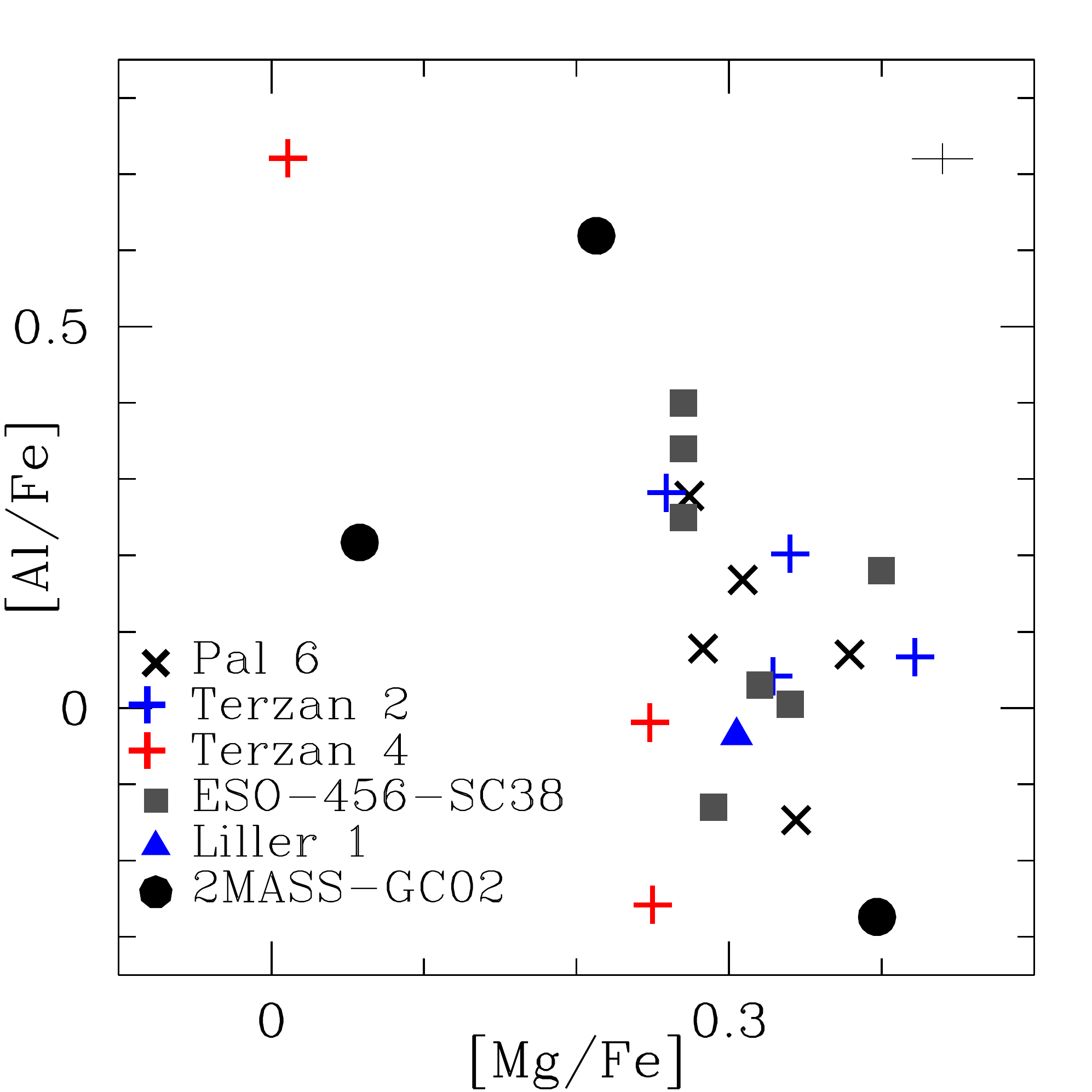}}}
{\subfigure{\includegraphics[height=4.25cm]{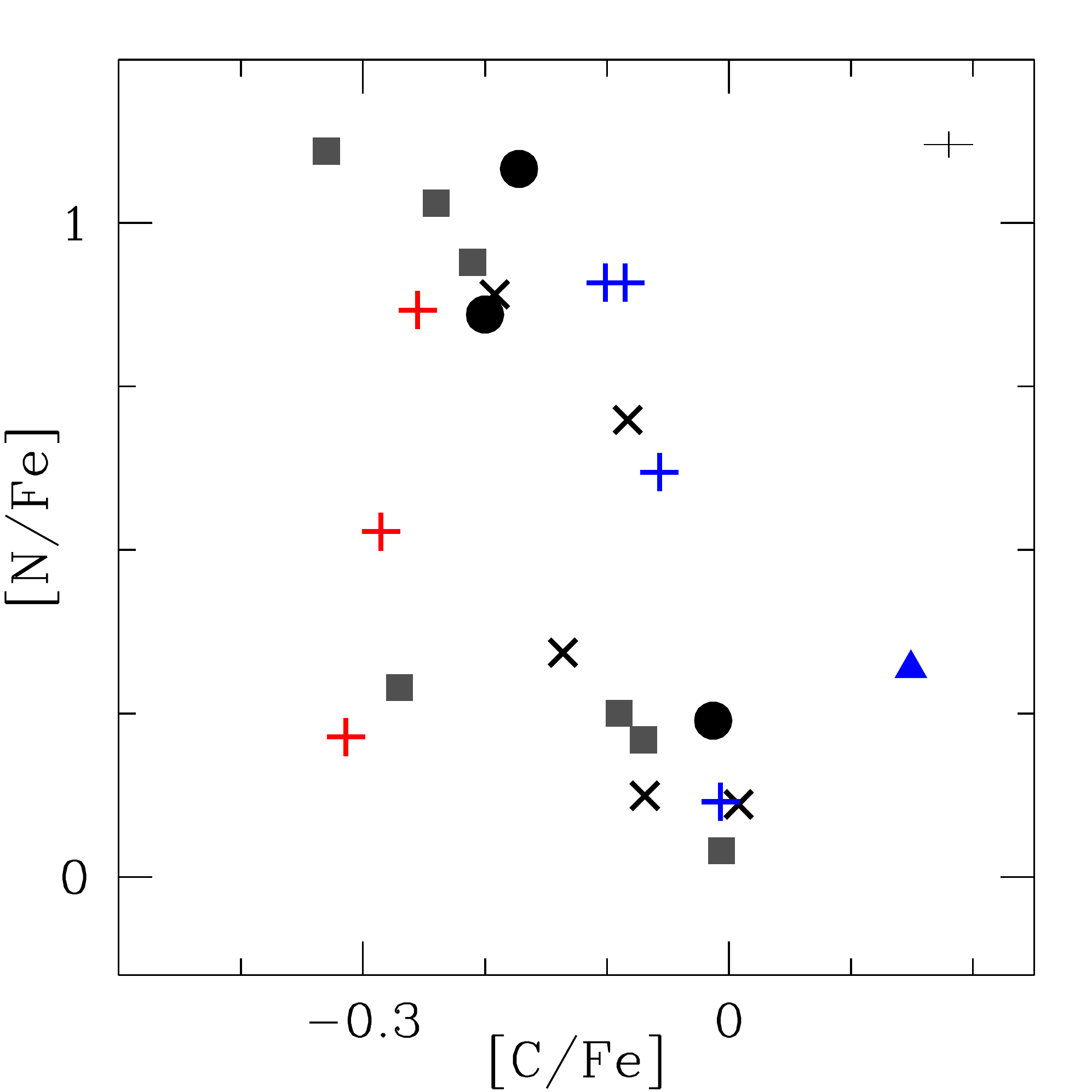}}}
{\subfigure{\includegraphics[height=4.25cm]{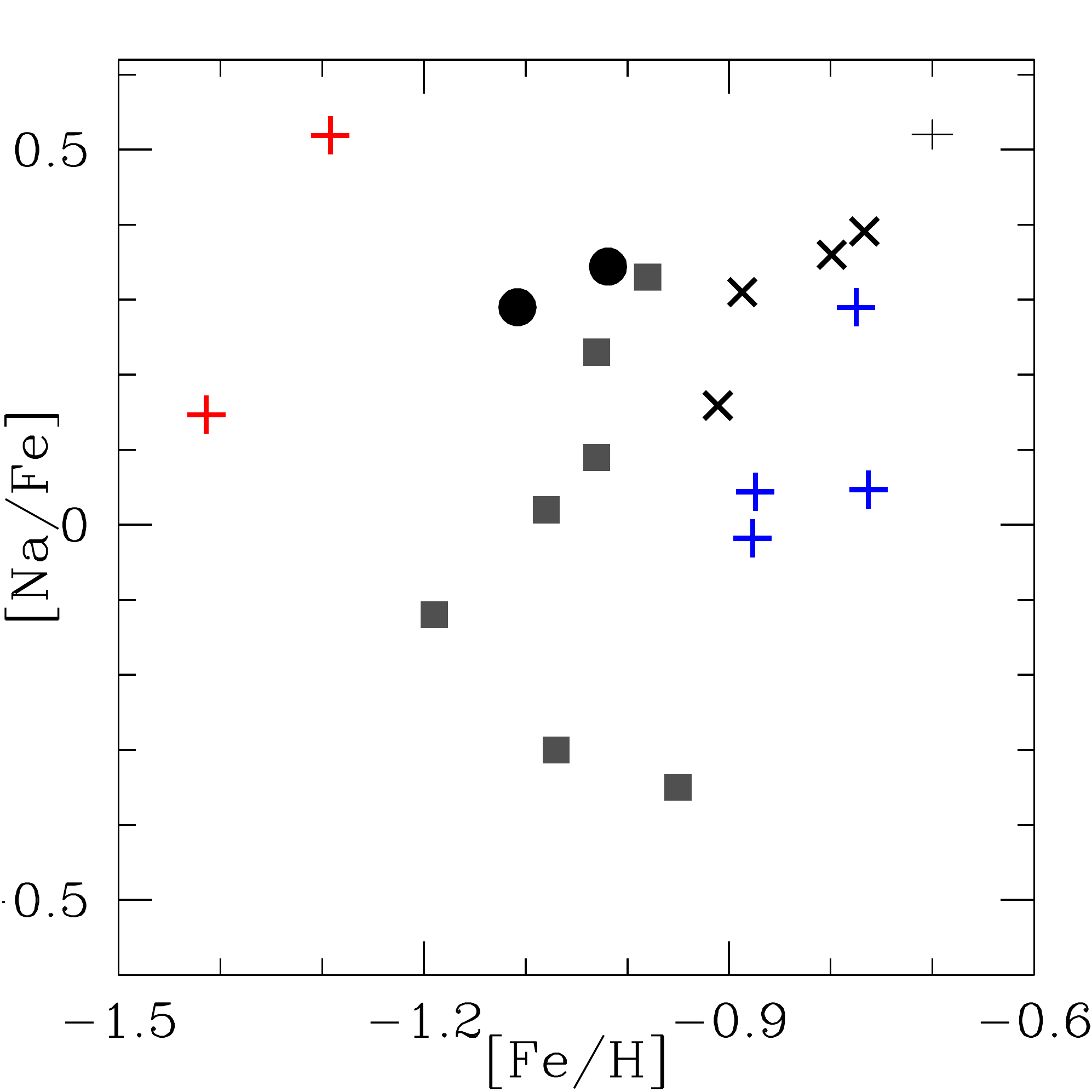}}}
\caption{The APOGEE chemical abundances for C, N, Na, Al and Mg in the 
cluster stars presented here.  The formal uncertainty in the APOGEE elemental abundances 
is indicated in the top-right.  All clusters show a wide spread of elemental abundances, 
indicating multiple populations co-exist within these bulge clusters.} 
\label{GC_abundances}
\end{figure}

Figure~\ref{GC_abundances} shows the elemental abundances of C, N, Na, Al and Mg of our 
candidate cluster stars.  In all these clusters (except for Liller~1 for which we have only one star 
with an APOGEE elemental abundance), a wide spread in elemental abundances is seen, 
indicative of multiple stellar populations.  Despite small number statistics for our clusters, 
the C-N anticorrelation is apparent as is the Mg-Al anticorrelation.
For example, we have only three stars in Terzan~4 with APOGEE elemental abundances, but their spread 
in $\rm [Al/Fe]$ ranges from $\sim-$0.25 to $\sim$+0.75~dex and 
their spread in $\rm [N/Fe]$ ranges from $\sim$0.2 to $\sim$0.8~dex.  
For all clusters studied here, stars with enhanced $\rm [N/Fe]$ abundances populate the cluster.
These peculiar chemical patterns appear to be ubiquitous for 
almost all GCs studied properly, and are also typical of stars in bulge 
clusters \citep[e.g.,][]{schiavon17, nataf19, fernandez19, fernandez20a}.  %

Some of the cluster stars are considerably aluminum enhanced, with $\rm [Al/Fe] >$+0.5, 
but still exhibit low levels of magnesium (e.g., $\rm [Mg/Fe] <$ 0.2).  This abundance pattern is similar 
to the $\sim$30 chemical peculiar APOGEE field stars reported toward the bulge region 
by \citet{fernandez20b}.  Their speculation that these aluminum enhanced metal-poor 
stars originated from GCs and were dynamically ejected into the bulge is supported by 
our observations of aluminum enhanced stars in the inner bulge GCs.
\begin{figure*}
\centering
\mbox{\subfigure{\includegraphics[height=8.5cm]{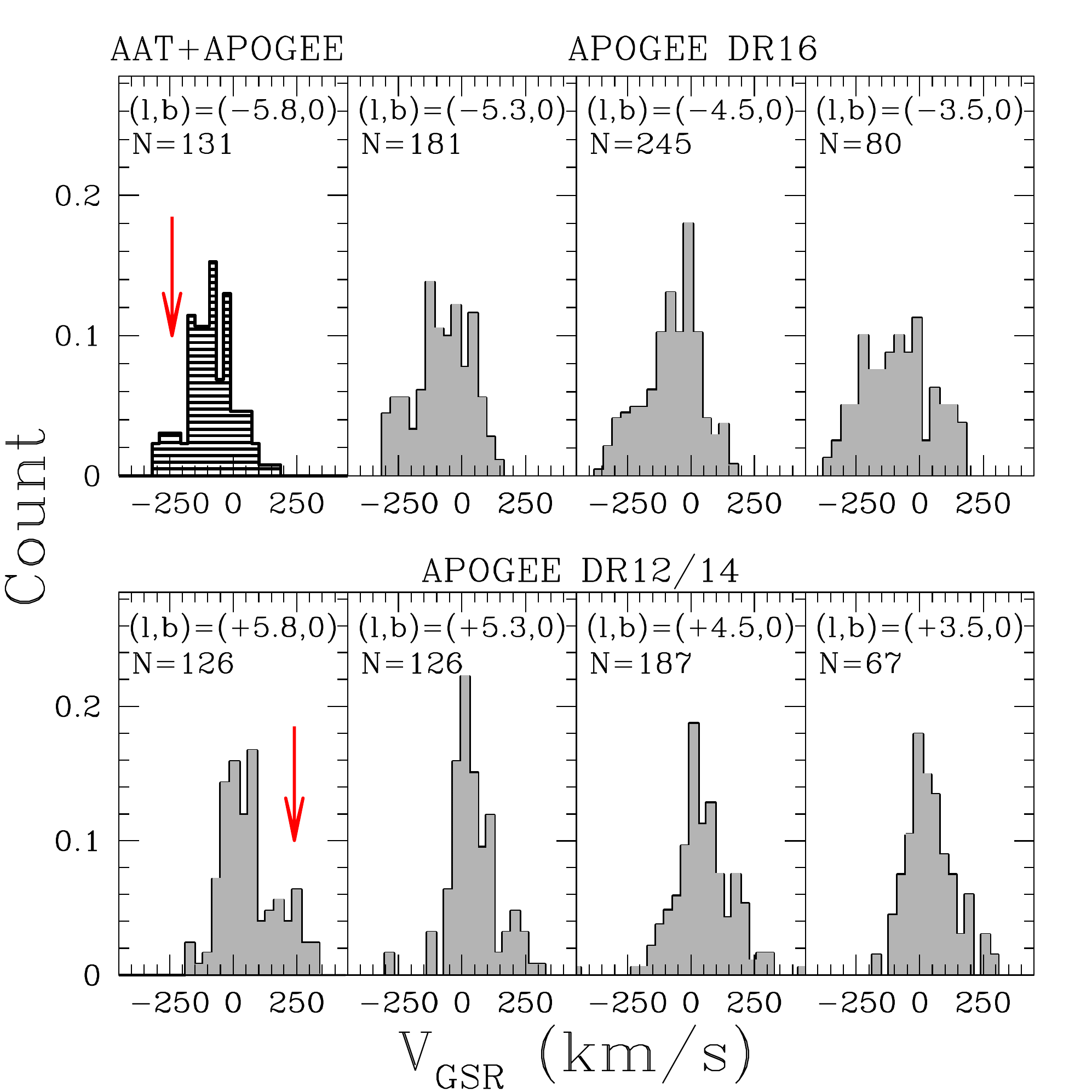}}}
{\subfigure{\includegraphics[height=8.5cm]{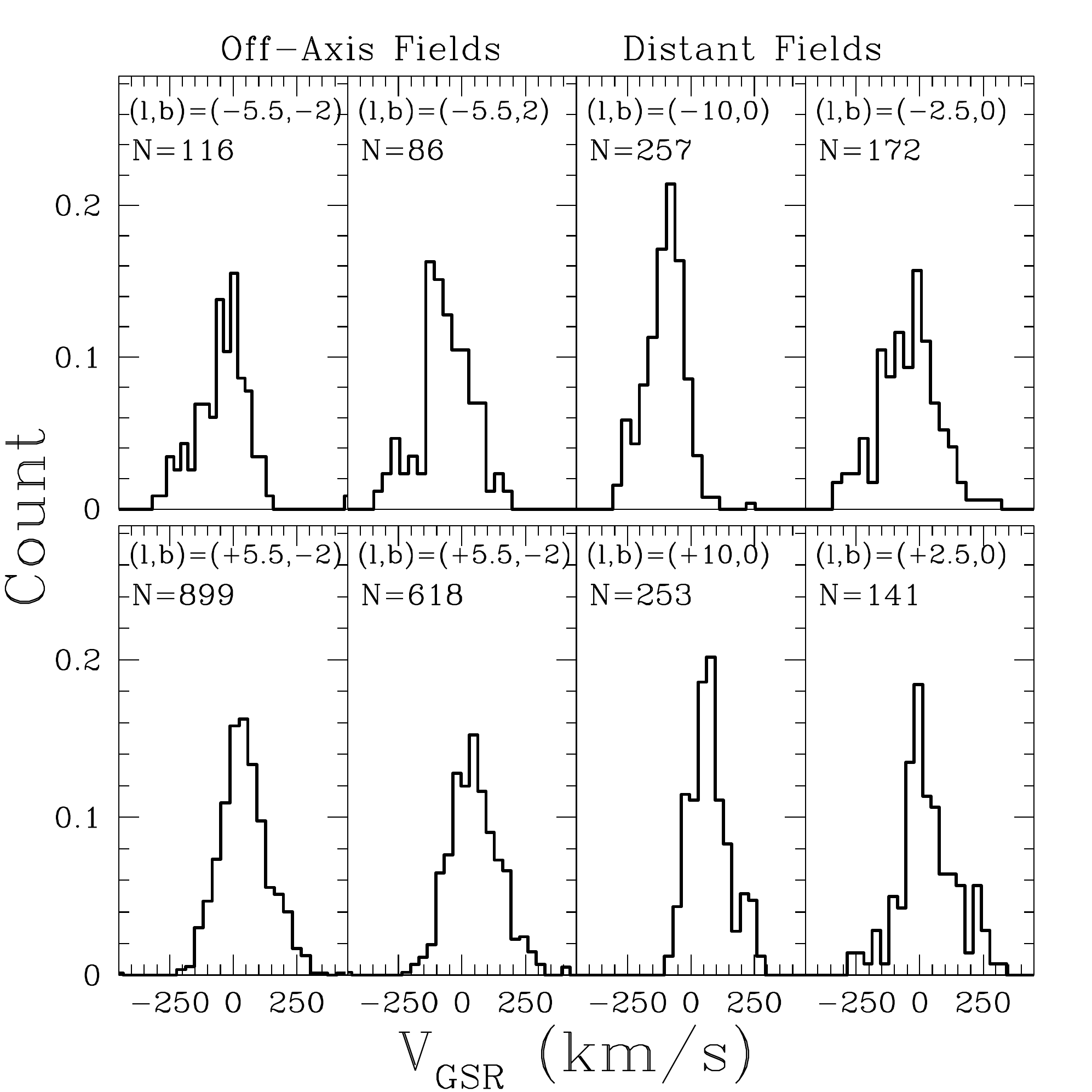}}}
\caption{The Galactocentric velocities of the APOGEE DR16 observations along the plane 
at negative longitudes along the plane (top) are compared to the APOGEE DR12/DR14 observations 
at positive longitudes (bottom).  For the ($l$,$b$)=($-$5.8,0) field, our new AAT observations have 
been folded in.  No strong $\sim$$-$200~km~s$^{-1}$ counter-feature to the ($l$,$b$)=(5.8,0) field is apparent.  
The arrows indicate where the expected high-$V_{GSR}$ would be if these stars were 
part of a kiloparsec-sized nuclear disk.  
{\it Right:} The Galactocentric velocities of the APOGEE observations in off-axis 
fields and along the plane where there should be no apparent nuclear feature.
%
}
\label{apogee_flds}
\end{figure*}
%
%
\section{The 200 km~s$^{-1}$ peak}
As discussed in the introduction, \citet{nidever12} identified cold velocity peaks in the 
radial velocity distribution of stars within $\sim$20$^\circ$ of the Galactic Center.  
Many of these peaks were shown to not be statistically significant \citep{zhou17} and there 
has also been the suggestion that APOGEE's selection function was biased \citep{aumer15}. 
Other studies have shown that some of these velocity peaks or shoulders can be associated 
with resonant bar orbits \citep{molloy15, mcgough20}.  It has also been suggested that the 
most prominent velocity peak at ($l$,$b$)=(6$^\circ$,0$^\circ$) is caused by stars moving 
on $\rm x_2$ orbits in a kiloparsec-sized nuclear disk or ring\citep{debattista15, debattista18}.

\subsection{Velocity Distributions}
When \citet{nidever12} first reported the so-called 200~km~s$^{-1}$ peak at positive longitudes along 
the plane of the Galaxy, there were no observations 
at the negative longitudes to search for a counter feature (see Figure~\ref{footprint}).  
This was unfortunate, because its prominence (or ``peakiness") and how wide spread it is at negative 
longitudes, could help discriminate between scenarios which attempt to explain this feature.  
For example, if the counter feature is wide-spread and appears as a shoulder instead of a 
discrete peak, this would agree with the models put forth 
by $e.g.,$ \citet{li14}, \citet{aumer15}, \citet{gomez16} and \citet{zhou20}, that the 200~km~s$^{-1}$ 
peak is composed of stars on 
bar-supporting orbits.  If the counter feature is confined to longitudes between $l \sim-$6$^\circ$ 
and $l \sim-$8$^\circ$ and is a discrete peak, this would agree with the suggestion 
of \citet{debattista18} that the 
200~km~s$^{-1}$ peak is composed of stars belonging to a nuclear disk or ring.

Figure~\ref{apogee_flds} shows the galactocentric velocities for the APOGEE DR16 at 
negative longitudes compared to those from APOGEE DR12/DR14 for the 
positive longitudes.  
Our new observations described in \S2 are also folded in, as they 
provide a more direct comparison to the ($l$,$b$)=(6$^\circ$, 0$^\circ$) field than using 
the APOGEE DR16 observations alone, as there are no APOGEE stars at longitudes 
between +6$^\circ$ and +7$^\circ$.  

A nuclear feature would be present mainly within $b\pm$1$^\circ$ and 
also mainly around $|l|\sim$5$^\circ$-8$^\circ$, as this is the line of sight 
tangent to the nuclear disk/ring \citep[$e.g.,$ see Figure~1;][]{debattista15}.  
The left panel of Figure~\ref{apogee_flds} concentrates 
on this region of the bulge.  The right panel shows the velocity distributions 
for stars at longitudes reaching to $|l|=$+10$^\circ$ and $|l|=2.5^\circ$ as well 
as fields $\pm$2$^\circ$ from the plane of the Galaxy.  Any high-$V_{GSR}$ features/signatures 
in these fields would not be due to a nuclear feature in these off-plane fields.  

A potential counter-peak at $\sim-$200~km~s$^{-1}$ 
appears to become more prominent and more separated from the rest of the distribution as the 
observations approach $|l|$ = 6$^\circ$, but the $l\sim$+6$^\circ$ observations show a significantly more 
pronounced peak as compared to at $l\sim-$6$^\circ$. 
Our radial velocity distributions show that stars do occupy the high-$V_{GSR}$ regime 
also around ($l$,$b$)=($-$5.5$^\circ$,0$^\circ$), but there is no obvious negative high-$V_{GSR}$ peak.  

\begin{figure}
\centering
\mbox{\subfigure{\includegraphics[height=5.0cm]{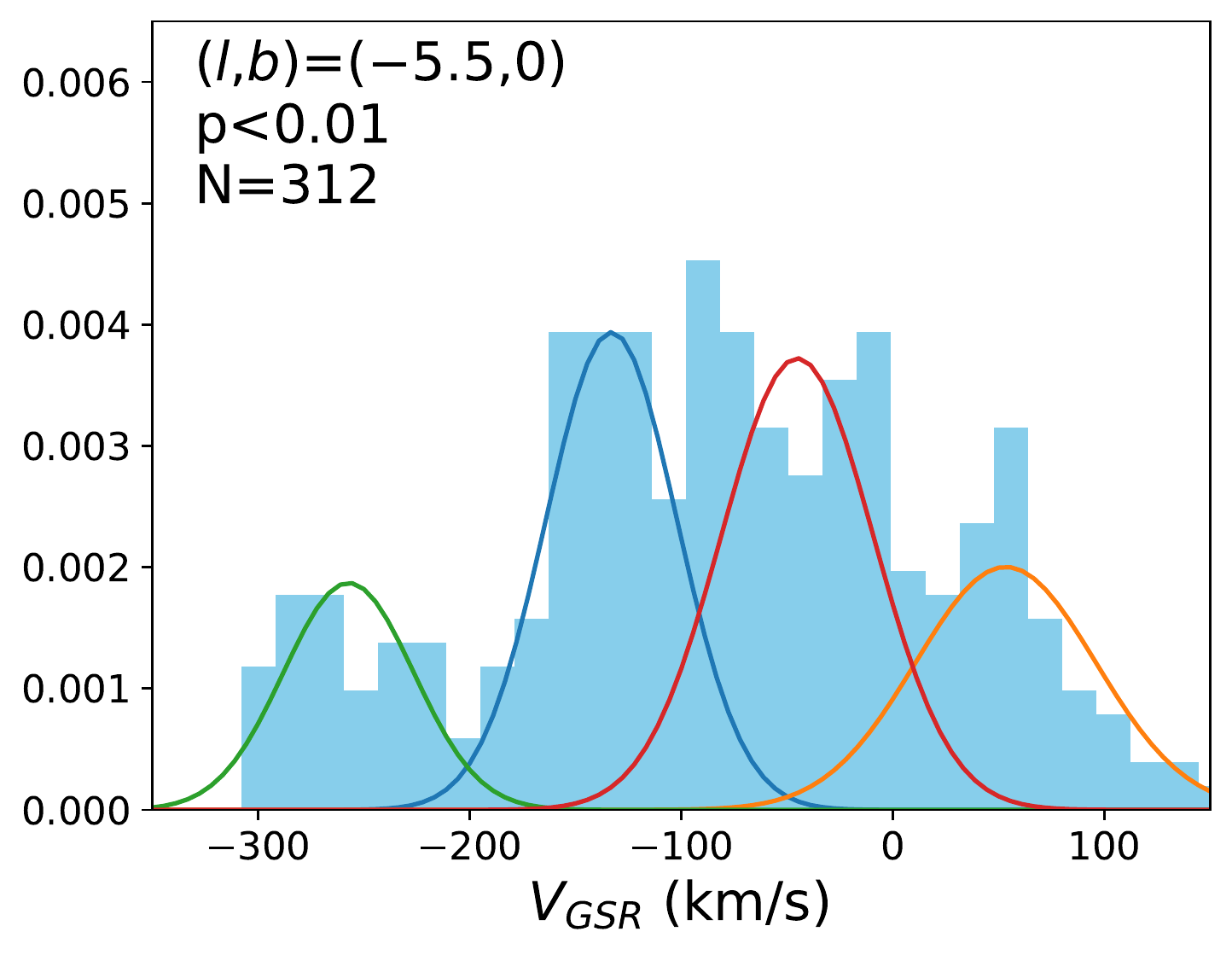}}}
{\subfigure{\includegraphics[height=5.0cm]{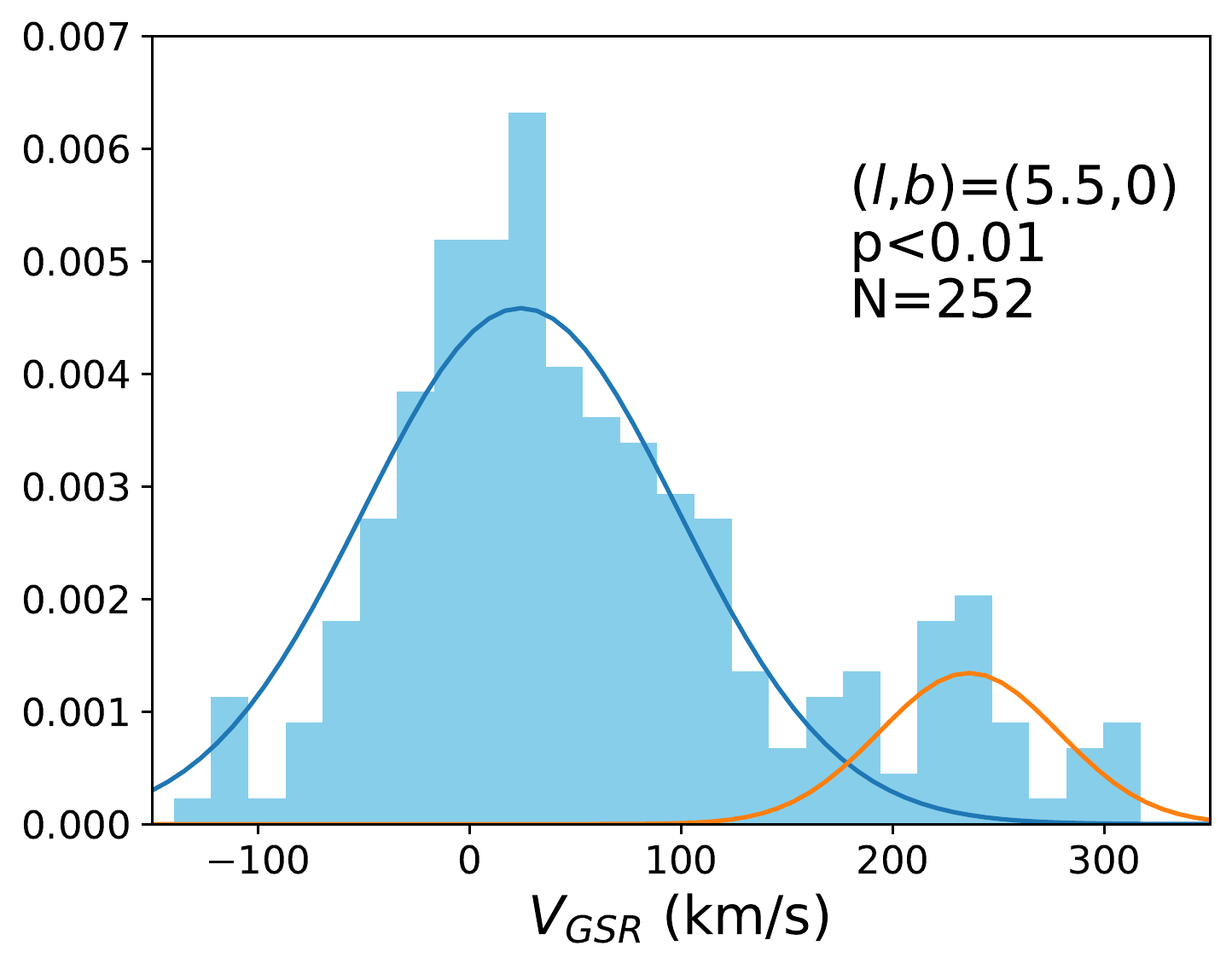}}}
\caption{A histogram of the Galactocentric velocities of APOGEE stars at 
($l$,$b$)=(5.5$^\circ$,0$^\circ$) (bottom)
and the APOGEE+AAT stars at ($l$,$b$)=($-$5.5$^\circ$,0$^\circ$) (top).  The best fit curve to the histogram 
from a $k$-means clustering 
algorithm and a GMM is overlaid.  The mean and standard deviations of these distributions are given 
in Table~\ref{stats}.  
}
\label{6degfields}
\end{figure}

Figure~\ref{6degfields} shows the APOGEE DR16 combined with the AAT velocities for 312 
stars in the field centered on ($l$,$b$)=($-$5.5$^\circ$, 0$^\circ$).  A Shapiro-Wilk test for 
normality on the radial velocity distribution gives $p$-values $<$ 0.05, indicating that the 
radial velocity distribution formally deviates from a Gaussian distribution.  A $k$-means 
clustering (KMC) algorithm is used to determine where the radial velocity distribution is 
aggregated.  A 4 peak curve provides the best fit to the ($l$,$b$)=($-$5.5$^\circ$, 0$^\circ$) 
velocity distribution, and a gaussian mixture model (GMM) indicates that the radial velocity 
peaks are at [$-$45~km~s$^{-1}$, $-$133~km~s$^{-1}$, 53~km~s$^{-1}$, $-$257~km~s$^{-1}$] 
with dispersions of [36~km~s$^{-1}$, 31~km~s$^{-1}$, 43~km~s$^{-1}$, 31~km~s$^{-1}$] 
and weights of [0.34, 0.31, 0.21, 0.14], respectively.  

This can be compared to the velocities of 252 stars in the field centered on 
($l$,$b$)=(5.5$^\circ$,0$^\circ$).  In this field, a 2 peak curve provides the best fit to the 
velocity distribution, and a gaussian mixture model (GMM) indicates that the radial velocity 
peaks are at [24~km~s$^{-1}$, 236~km~s$^{-1}$] with dispersions 
of [75~km~s$^{-1}$, 43~km~s$^{-1}$] and weights of [0.86, 0.14], respectively.  Whereas 
the ($l$,$b$)=($-$5.5$^\circ$, 0$^\circ$) shows a 
complicated distribution of velocities, the ($l$,$b$)=(5.5$^\circ$,0$^\circ$) field exhibits a clear 
double peak, with one of the main peaks being at a positive high-$V_{GSR}$.  That the bulge velocity 
distribution is complex is not surprising given the large 
velocity dispersions reported especially in bulge fields close to the Galactic 
plane \citep[e.g.,][]{kunder12}, and that the bulge is made up of a complex variety of 
stellar populations coexisting in the central regions of the Milky 
Way \citep[e.g.,][and references therein]{recioblanco17, queiroz20, rojasarriagada20}.

Table~\ref{stats} lists the Shapiro-Wilk $p$-value, the number of peaks computed from $k$-means, 
and the GMM values of where the peaks are located for the fields shown in Figure~\ref{6degfields}.
Although formally these statistical tests provide a quantitative characterization of the 
velocity distribution of stars near the plane of the Galaxy, we believe 
the large velocity dispersion combined with the sample size makes it difficult to conclusively 
determine the underlying velocity distribution.  

It is evident that any preference for an excess of stars at $V_{GSR}\sim$$-$200~km~s$^{-1}$ 
in the $l=-$5.5$^\circ$ field is not as extreme as was seen in the $l=$+5.5$^\circ$ field, 
nor is it confined to only this field.  For example, the field at ($l$,$b$)=($-$10$^\circ$, 0$^\circ$) and 
at ($l$,$b$)=($-$4$^\circ$, 0$^\circ$) also may show some indication for 
a $V_{GSR}\sim$$-$200~km~s$^{-1}$ peak.

The Figure~\ref{apogee_flds} presented here is similar to the velocity distributions shown 
in \citet[][their Figure~2]{zhou20}, except that we have $\sim$100 more stars with radial 
velocities centered on ($l$,$b$)=($-$6$^\circ$,0$^\circ$) due to our new AAT observations.
We also show velocity histograms spanning ranges of longitude to specifically explore a counter-feature 
at negative longitudes.  Instead of selecting 2~degree fields based on where the positive longitude fields are 
centered, we separate the stars so they naturally follow the sparser coverage along the 
negative longitudes (see Figure~\ref{footprint}).  In particular, as there is no negative longitude field 
centered on ($l$,$b$)=($-$6$^\circ$,0$^\circ$), we do not show a velocity distribution for this field.  Instead, 
the ($l$,$b$)=($-$5.8$^\circ$,0$^\circ$) field presented here 
spans half a degree from $-$5.5$^\circ$ to $-$6.0$^\circ$, and 
the ($l$,$b$)=($-$5.3$^\circ$,0$^\circ$) field spans half a degree from $-$5.0$^\circ$ to $-$5.5$^\circ$.
These fields can 
be directly compared to the APOGEE stars at positive longitudes and hence more closely 
mirrors the APOGEE observations in hand.  

\subsection{Velocity and Distance Correlations}
\begin{figure}
\centering
\mbox{\subfigure{\includegraphics[height=6.25cm]{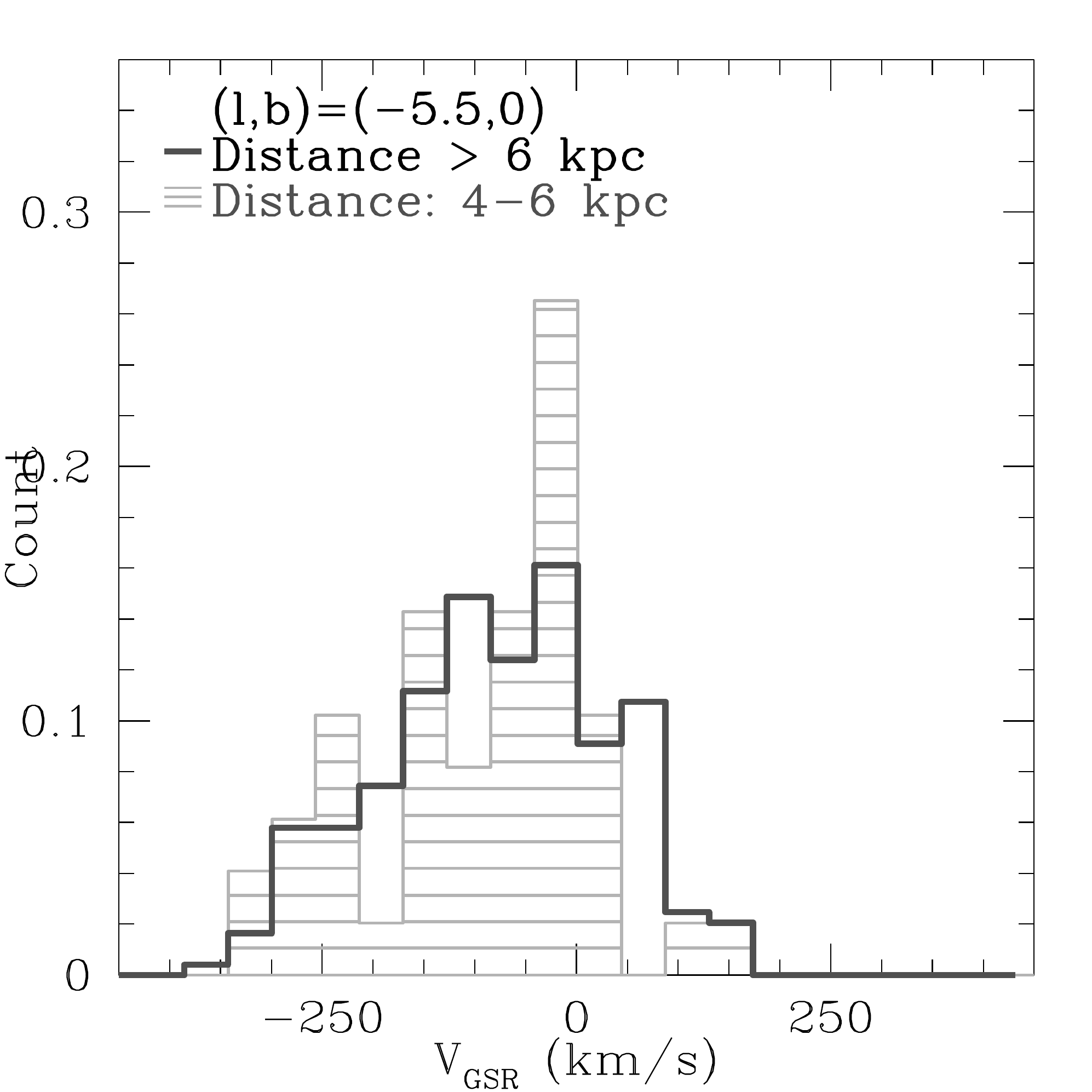}}}
{\subfigure{\includegraphics[height=6.25cm]{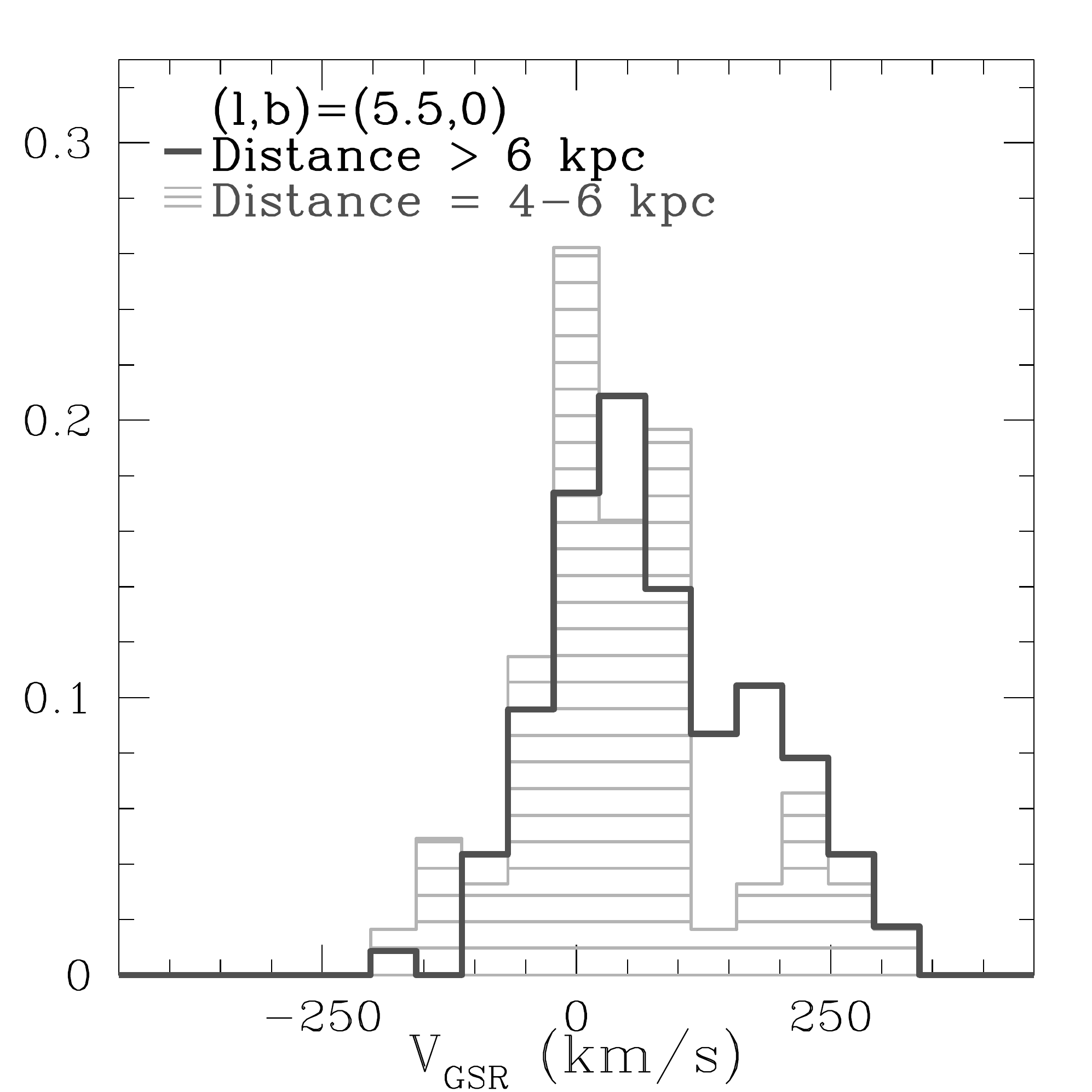}}}
\caption{The Galactocentric velocity distribution of APOGEE Stars $-$6\degree~field (top) 
and the +6\degree~field (bottom) separated by heliocentric distance.  In general, stars with larger distances 
(solid line) have larger velocities.  }  
\label{AP_Distance}
\end{figure}

\citet{li14} and \citet{zhou17} use Milky Way simulations to remark that  
in a distance-velocity diagram, higher velocity particles 
should be at larger distances.  This is due to the geometric intersections between the line-of-sight and the 
particles orbit;  at the tangent points of a star's orbit, it will have the highest velocity and remain 
there longer relative to us.  The StarHorse distances \citep{queiroz20} 
released for the APOGEE DR16 stars can be used to explore if the high velocity stars seen in the 
histograms presented above are consistent with particle motions in an axisymmetric disk.
  
Figure~\ref{AP_Distance} shows the velocity distributions for those stars with StarHorse distances 
between 4-6~kpc compared to the velocity distributions for those stars with distances $>$6~kpc.  
In both fields there are still stars with distances between 4-6~kpc that occupy the high-$V_{GSR}$ 
peaks.  There is only a mild trend for stars with on average larger distances to have 
larger velocities.  The stars with distances 
between 4-6~kpc have dominant velocity peaks at $\sim |$50$|$~km~s$^{-1}$.  In contrast, 
stars with distances larger than 6~kpc are not as strongly peaked and instead have tails extending 
to larger velocities.  This suggests that the high-$V_{GSR}$ peak 
can not be explained by stars at larger distances alone, although it may also be that the 
StarHorse distances are not reliable for these stars.  The plane of the Galaxy has especially high 
and variable extinction values and this may affect the accuracy of the StarHorse distances.

\begin{figure*}
\centering
\mbox{\subfigure{\includegraphics[height=8.0cm]{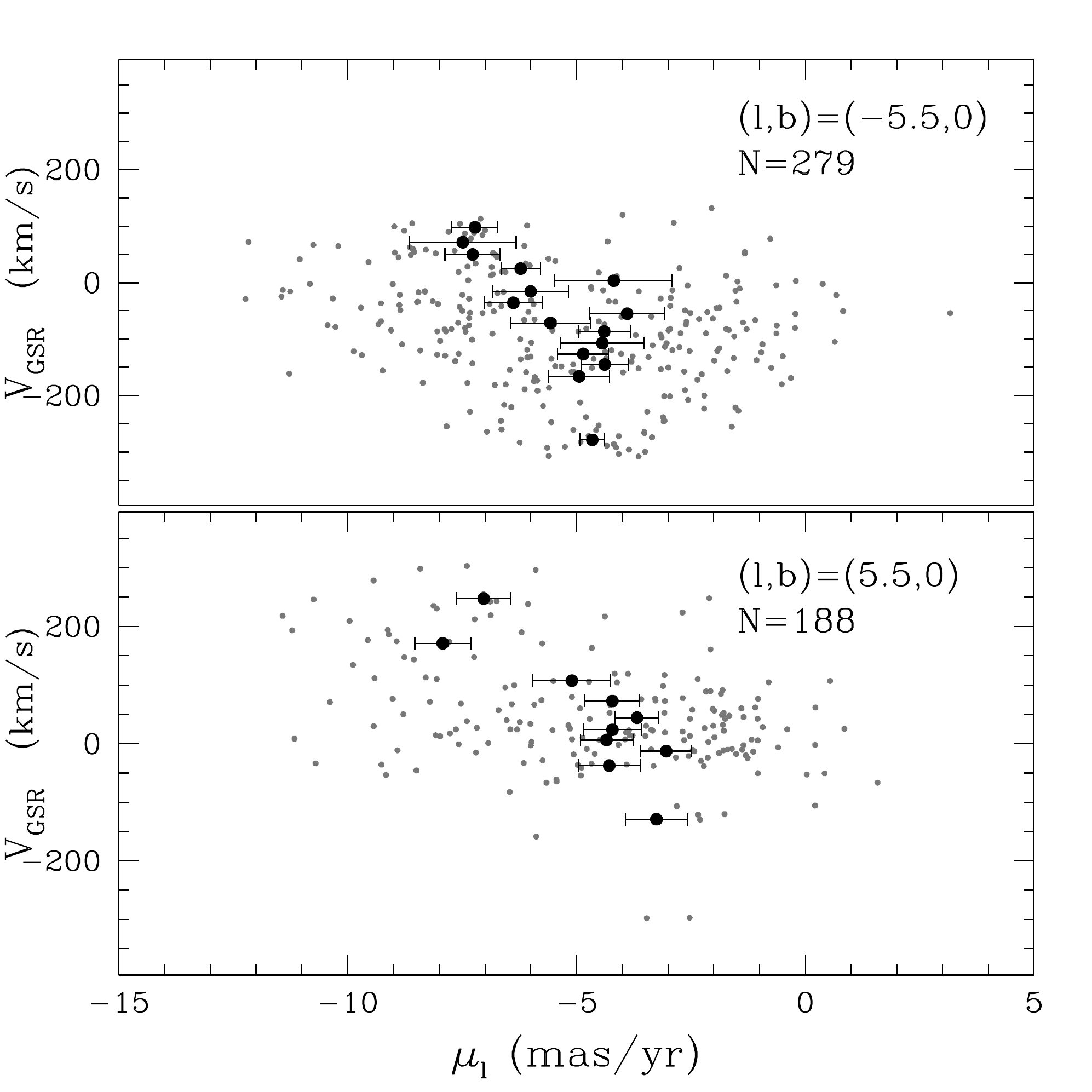}}}
{\subfigure{\includegraphics[height=8.0cm]{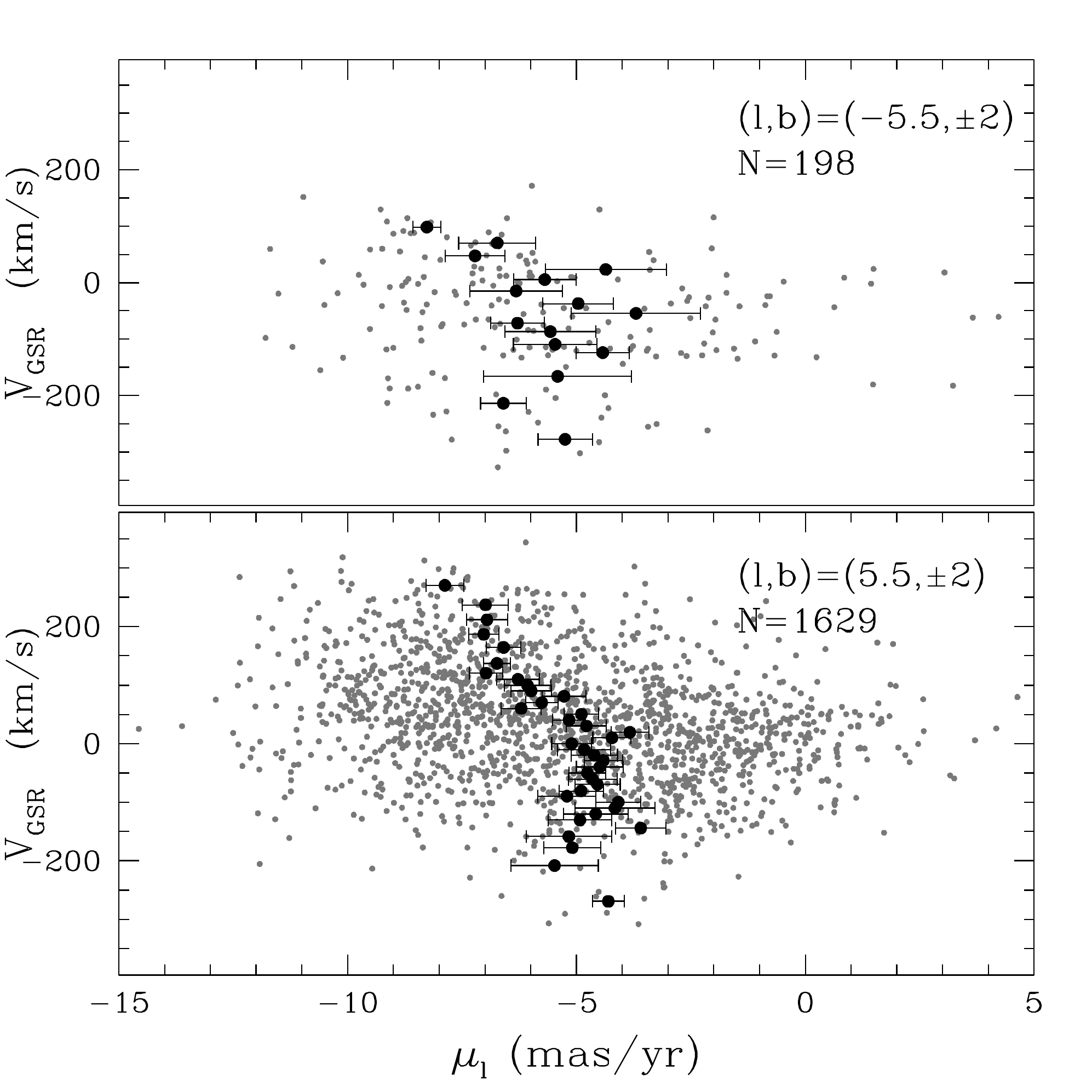}}}
\caption{The $V_{GSR}$ velocities as a function of proper motion,  $\mu_\ell$, for the fields at positive (top) 
and negative (bottom) longitudes.  Whereas the positive high-$V_{GSR}$ stars have proper motions 
offset from the rest of the bulge, this is not the case for the negative high-$V_{GSR}$ stars.} 
\label{pms200}
\end{figure*}
\subsection{Velocity and Proper Motion Correlations}
{\it Gaia} proper motions are available for a number of stars along the plane, and in \S3 
we showed that they can be used also in this crowded region of the Galaxy.  In particular, 
from an analysis of globular cluster stars, a star-to-star scatter of 0.15~mas~yr$^{-1}$ or less 
can be expected.  Formally the precision of the EDR3 proper motions is 
$\sim$0.07 mas~yr$^{-1}$ (at $G$=17) for {\it Gaia} EDR3 as compared to 
0.2 mas~yr$^{-1}$ (at $G$=17) in {\it Gaia} DR2 \citep{lindegren20}.  
{\it Gaia} EDR3 proper motions offer not only an increase in precision, but the number of stars with proper 
motions along the plane also increases by $\sim$15\% as compared to {\it Gaia} DR2.  Therefore, our 
sample of stars along the plane is useful to probe the correlation between the high-$V_{GSR}$ 
regime also in proper-motion space.

Figure~\ref{pms200} shows the correlation between velocity and $\mu_\ell$ for 249 stars 
at ($l$,$b$)=(5.5$^\circ$,0$^\circ$).  At a velocity of $V_{GSR} >$100 km~s$^{-1}$, 
stars have proper motions with larger $|\mu_\ell|$ values than the rest of the sample.  
In particular, stars with $>$100~km~s$^{-1}$ have $\mu_\ell$ of $\sim$ $-$7.5 mas~yr$^{-1}$ 
as compared to $\mu_\ell$ of $-$4 mas~yr$^{-1}$ for the bulge field.

The $\mu_\ell$ for 328 stars at ($l$,$b$)=($-$5.5$^\circ$,0$^\circ$) shows the same trend for 
the positive high-$V_{GSR}$ stars -- the stars with $\sim$200~km~s$^{-1}$ velocities 
have $|\mu_\ell|$ proper motions larger than the rest of the sample.  In contrast, the 
negative high-$V_{GSR}$ stars have a similar $\mu_\ell$ to the bulge field.  There is no 
preponderance for the negative high-$V_{GSR}$ regime to have different $\mu_\ell$ proper 
motions to the bulge field.  

The same trend in proper motion is also seen $\pm$2$^\circ$ from the plane.  There is no 
indication that the stars along the plane of the Galaxy at $|l|=$5.5$^\circ$ differ 
in $\mu_\ell$ as compared to those $\pm$2$^\circ$ from the plane at $|l|=$5.5$^\circ$.

\citet{mcgough20} use 71 stars at $l\sim$6$^{\circ}$, 52 stars at $l\sim$8$^{\circ}$ 
and 27 stars at $l\sim$4$^{\circ}$ to show that the high-$V_{GSR}$ 
stars occupy a smaller range in $\mu_\ell$ than the full sample of stars.  They attribute 
this to stars on periodic orbits 
in an exponential bar, such as propellor orbits.  They were not able to investigate 
the proper motions at negative longitudes.  \citet{zhou20} corroborate the result from 
\citet{mcgough20} result using DR2 proper motions for the APOGEE DR16 sample.  
Here we are able to offer additional insights on the correlation between velocity and $\mu_\ell$ in the 
bulge using the better precision and sampling of the EDR3 proper motions combined 
with both the APOGEE and our new AAT observations.  


Besides {\it Gaia} proper motions, proper motions of bulge stars from \citet{smith18} are publicly 
available through the VIRAC catalog, 
a near-infrared proper motion and parallax catalogue of the VISTA Variables in the Via Lactea (VVV) 
survey.  Out of the 3614 APOGEE stars within $|b| <$ 1$^\circ$, $(J-K)_0 >$ 0.5 and 
log~$g >$ 3.8, about half of those (1656) have a VIRAC proper motion, and about a third 
of those (592) have a {\tt flag}=1, indicating it is reliable.  In particular, in the ($-$5.5,0) field, 
there are 220 stars with VIRAC proper motions, and in the (+5.5,0) field there are 98 stars 
with VIRAC proper motions.  This overlap is smaller then when using {\it Gaia} EDR3.  
There are 1510 APOGEE stars within $|b| <$ 1$^\circ$, $(J-K)_0 >$ 0.5 and 
log~$g >$ 3.8 that have a {\it Gaia} EDR3 proper motion; 87\% of these have absolute proper motion 
uncertainties $<$1~mas~yr$^{-1}$.  The number of APOGEE stars with {\it Gaia} proper motions 
in the ($-$5.5,0) and (5.5,0) field is greater by a factor of 1.5 - 2.  

Using the VIRAC proper motions results in the same trends as seen above in 
Figure~\ref{pms200} using the more extensive {\it Gaia} EDR3 proper motions.  
APOGEE stars with radial velocities of $\sim$200 km~s$^{-1}$ have 
larger $|\mu_l|$ values.  This is the case both for the bulge stars with $|b|<1^\circ$ as 
well as the stars at $|b|\sim 2^\circ$.

We prefer not to combine the {\it Gaia} and VIRAC proper motion catalogs due to the offsets 
between these catalogues caused by the drift motion of the pool of reference stars used for 
each pawprint set in VIRAC \citep[see e.g.,][]{smith18, clarke19}.  These are estimated to 
be +2.20 mas~yr$^{-1}$ in $\mu_\alpha$ and $-$4.85 mas~yr$^{-1}$ in $\mu_\delta$; 
the authors mention it is quite probable that there 
are unknown systematic uncertainties besides these offsets as well. 

\section{Discussion}
Proper motions from {\it Gaia} EDR3, radial velocities from APOGEE, distances from StarHorse, 
and our own observations from the AAT are combined for stars along the plane of the Galaxy 
with the aim of testing the hypothesis that the high-$V_{GSR}$ stars are the signature 
of a nuclear disk or ring \citep{debattista18}.  Our new observations from the AAT 
at ($l$,$b$)=($-$6$^\circ$,0$^\circ$) are combined with the APOGEE DR16 observations 
to increase the sample size of stars with radial velocities at negative longitudes where the tangent 
point of a nuclear feature should be evident.  

From a GMM analysis the putative negative high-$V_{GSR}$ peak at 
($l$,$b$)=($-$5.5$^\circ$,0$^\circ$) is at $V_{GSR}$$\sim$$-$250~km~s$^{-1}$ 
whereas
the positive high-$V_{GSR}$ peak at ($l$,$b$)=(5.5$^\circ$,0$^\circ$) 
is at $V_{GSR}$$\sim$230~km~s$^{-1}$. 
Therefore, the negative high-$V_{GSR}$ stars have the larger $|V_{GSR}|$,
in contrast to what is expected for a nuclear feature.

Along the plane of the Galaxy, the new {\it Gaia} EDR3 proper motions increase the number 
of APOGEE stars with proper motions by $\sim$15\% as compared to {\it Gaia} DR2.  
The proper motions of the high-$V_{GSR}$ stars occupy a small range in proper motions, 
spanning $\sim$5 mas~yr$^{-1}$.  This is inconsistent with what is predicted for stars in a nuclear feature, 
as proper motions of stars in both a nuclear disk and nuclear ring would span 
a wider range of  in $\mu_\ell$, $\sim$10~mas~yr$^{-1}$ \citep{debattista18}.

The $\mu_\ell$ of the positive high-$V_{GSR}$ exhibit larger 
$|\mu_\ell|$ proper motions than the bulge stars.  However, the negative high-$V_{GSR}$ 
stars have similar $\mu_\ell$ proper motions as bulge stars.  
\citet{zhou20} argues that this observed $V_{GSR}$ - $\mu_l$ distribution is 
consistent with the predictions of a simple MW bar model as presented in \citet{shen10}.  
This is especially the case if the APOGEE stars were to lie predominantly in front of the 
Galactic Center.  However, the \citet{shen10} model does not predict a distinct peak 
at ($l$,$b$)=(5.5\degree,0\degree).  The only way to produce this peak in the 
\citet{shen10} model is if the high-$V_{GSR}$ stars are at distances at 
$\sim$8.5$\pm$1~kpc \citep[see e.g., Fig.~3 in][]{li14}.  Accordingly there is 
a tension in the  \citet{shen10} model between the APOGEE stars needing to be at $\sim$8.5~kpc 
to explain any high-$V_{GSR}$ peak but to be primarily in front of the Galactic center to match 
the $\mu_l$ proper motions.  

Although there are negative high-$V_{GSR}$ stars residing in the negative longitude fields, the 
negative high-$V_{GSR}$ peak is not nearly as prominent or discrete as seen in the 
positive high-$V_{GSR}$ peak at the positive longitudes (see Figure~\ref{apogee_flds}).
Stars on propellor orbits as described by \citet{mcgough20} are not able to explain this.  
The propellor orbits can produce positive high-$V_{GSR}$ stars with similar proper motions 
to what is observed, but there is no reason why APOGEE would select just these orbits 
at ($l$,$b$)=(5.5\degree,0\degree).
Although it has been hypothesized that the APOGEE selection function 
preferably selects young stars \citep{aumer15}, stars on propellor orbits would not necessarily 
be young, and there further has been no observational evidence to suggest that high-$V_{GSR}$ stars are 
in fact younger than the bulk of the population\citep{zasowski16, zhou17, zhou20}.
Therefore, those propellor orbits would give rise to a strong secondary peak only at ($l$,$b$)=(5.5\degree,0\degree) 
is not a satisfactory way of explaining a distinct high-$V_{GSR}$ peak.

It was shown in \citet[e.g.,][]{li14} the high-$V_{GSR}$ stars can be explained by stars at larger 
distances, although this would produce a high-$V_{GSR}$ peak, instead of a shoulder.  
Using StarHorse distances we are not able to confirm that the high-$V_{GSR}$ stars are at further 
distances.  This leaves open a different interpretation for the peak in the APOGEE high-$V_{GSR}$ 
stars than current Milky Way models of the orbits of bulge stars suggest.  

Our analysis assumes that the completeness between APOGEE-2 and APOGEE for the bulge 
fields are similar without major differences in the selection effects.  This is the case for other studies 
using APOGEE DR16 to compare the 
positive and negative longitudes of the inner galaxy \citep[e.g.,][]{zhou20, rojasarriagada20}.  
\citet{zasowski17} reports that in both APOGEE-2 and APOGEE, bulge giants are selected based on their 
dereddened $\rm (J-K)$ colors, requiring $\rm (J-K)_0 >$0.5, and Figure~\ref{cmd} shows no obvious 
indication that the APOGEE stars 
at negative longitudes do not probe the full $\rm (J-K)_0$ color range for bulge giants.  
Although, there have been no detailed investigations to investigate selection effects 
in APOGEE DR16 \citep[see e.g.,][]{nandakumar17, fragkoudi18}.  \citet{queiroz21} uses cartesian 
density maps of the bulge stars to surmise that the complex mix of stellar populations and selection 
function of APOGEE does provide a homogeneous coverage of the entire inner Galaxy.  
Fortunately, the detection of a high-velocity peak at
negative longitudes should be as apparent, if not more, using stars at
negative longitudes of the Galaxy \citep{debattista18}.  Yet still, our different cuts and stellar
divisions within the APOGEE-2 dataset (e.g., Figure~\ref{apogee_flds}, Figure~\ref{6degfields}, Table~\ref{stats})
does not show any obvious $-$200 km~s$^{-1}$ peak.  

\section{Conclusions}
The APOGEE DR16 observations combined with {\it Gaia} EDR3 proper motions are used to 
investigate the hypothesis that the cold high-velocity peak 
\citep[at $V_{GSR} \sim \rm 200~km~s^{-1}$,][]{nidever12} are due to a kilo-parsec 
nuclear feature \citep{debattista15, debattista18}.  The APOGEE DR16 
data in the negative longitudes were examined for a counter-peak at $V_{GSR} \sim -$200~km~s$^{-1}$.  
Some evidence for a high velocity shoulder in a number of different 
APOGEE negative longitude fields along the plane is seen, but no 
dominant $V_{GSR} \sim-$200~km~s$^{-1}$ feature is apparent.  
Unfortunately the sample size is small compared to the large velocity dispersion of the 
bulge population along the plane of the Galaxy, and a larger sample size as well as wider 
coverage (e.g.,  $l= -$6$^\circ$ to $-$8$^\circ$) would offer a 
deeper view on characterizing any $\sim|200|$~km~s$^{-1}$ feature.  
Distances and proper motions of the high velocity stars are inconsistent with 
the $\sim|200|$~km~s$^{-1}$ stars belonging to different classes of bar orbits from 
model for galactic bars as put forth by previous investigations \citep[e.g.,][]{li14, zhou17, mcgough20, zhou20}.  
They are also inconsistent with the idea that the MW harbors a kilo-parsec 
nuclear feature \citep{debattista15, debattista18}.
We therefore currently have no single model that can explain the high-$V_{GSR}$ observations.

To corroborate the accuracy of the {\it Gaia} EDR3 proper motions along the plane of the Galaxy, 
APOGEE stars in a number of globular clusters are identified without the aid of proper motion criteria.
Many of these stars had previously not been reported; the first APOGEE stars in the globular clusters 
2MASS-GC02 and Terzan~4 are presented, both of which show evidence of multiple populations.  
The average radial velocity of the stars 
in 2MASS-GC02 is $-87 \pm 7$~km~s$^{-1}$, which is more than 150~km~s$^{-1}$ offset from 
the literature value \citep{borissova07}.

\acknowledgements
We thank the referee for suggestions that improved the quality of our analysis.
A.M.K and R.E.C. are thankful to colleagues Danielle Miller and Andrew Schulz for aiding with 
the AAT observations and for stimulating conversations.  AMK acknowledges support from grant 
AST-2009836 from the National Science Foundation.  The grant support provided, in part, by the 
M.J. Murdock Charitable Trust (NS-2017321) is acknowledged.
V.P.D. is supported by STFC Consolidated grant \#ST/R000786/1. 
A.J.K-H gratefully acknowledges funding by the Deutsche Forschungsgemeinschaft 
(DFG, German Research Foundation) -- Project-ID 138713538 -- SFB 881 (``The Milky Way System''), 
subprojects A03, A05, A11. 

\begin{deluxetable*}{cccccc}
\tablecaption{Targeted Globular Cluster Properties
\label{GCTab}}
\tabletypesize{\scriptsize}
\tablehead{
\colhead{GC} & \colhead{ $\mu_{\alpha}, \mu_{\delta}$ (This work)} & \colhead{Dist} &
\colhead{$v_r$} & \colhead{$\mu_{\alpha}, \mu_{\delta}$ \citep{rossi15}} & \colhead{$\mu_{\alpha}, \mu_{\delta}$  \citep{vasiliev21}}  \\
\colhead{} & \colhead{ (mas~yr$^{-1}$)} & \colhead{(kpc)} &
\colhead{(km~s$^{-1}$) } & \colhead{\rm (mas~yr$^{-1}$)} & \colhead{($\rm mas~yr^{-1}$)} 
\\
}
\hspace{-2cm}
\startdata
2MASS-GC02 & - & 7.1 & $-$87 & - & $-1.97\pm0.16$, $-3.72\pm0.15$  \\ 
Palomar6 & $-9.27\pm0.04$, $-5.34\pm0.13$ & 5.8 &181 & 2.95$\pm$0.41, 1.24$\pm$0.19 &  $-9.20\pm0.04$, $-5.32\pm0.03$\\ 
Terzan 2 & $-2.13\pm0.03$, $-6.34\pm0.06$ & 7.5 &109 & $0.94\pm0.30$, 0.15$\pm$0.42 &  $-2.17\pm0.04$, $-6.25\pm0.04$  \\
Terzan 4 & $-5.21\pm0.15$, $-3.61\pm0.07$ & 7.2 &$-$50 & 3.50$\pm$0.69, 0.35$\pm$0.58 & $-5.62\pm0.07$, $-3.62\pm0.07$ \\
Liller 1 & $-5.34\pm0.14$, $-7.37\pm0.12$ & 8.1 &52 & - & $-5.40\pm0.13$, $-7.48\pm0.10$ \\
ESO 456-SC38* & 0.73$\pm$0.04, $-$3.06$\pm$0.07 & 8.8 & $-$150 & 3.08$\pm$0.29, 2.00$\pm$0.34 & 0.67$\pm$0.04, $-$2.99$\pm$0.03 \\ 
\enddata
\tablecomments{$^*$from \citet{kunder20b}, updated EDR3 proper motions}
\end{deluxetable*}
%
%

%
\begin{table*}
\begin{scriptsize}
\centering
\caption{Parameters and Abundances of Globular Cluster Stars
\label{chartable}}
\begin{tabular}{p{1.1in}p{0.35in}p{0.35in}p{0.35in}p{0.4in}p{0.4in}p{0.35in}p{0.35in}p{0.35in}p{0.35in}p{0.35in}p{0.35in}} \\ \hline
APOGEE ID & RV & r & Distance & $\mu_{\alpha}$ & $\mu_{\delta}$ & $\rm [Fe/H]$ & $\rm [C/Fe]$ & $\rm [N/Fe]$ & $\rm [Na/Fe]$ & 
$\rm [Mg/Fe]$ & $\rm [Al/Fe]$ \\ \hline
\textbf{2MASS-GC02} & & & & & & & & & & \\
2M18092967-2048019 & $-$83.018 & 0.442 & \nodata & \nodata & \nodata & $-$1.108 & $-$0.013 & 0.239 & 0.290 & 0.397 & $-$0.274 \\
2M18092981-2045283 & $-$96.486 & 0.305 & \nodata & \nodata & \nodata & \nodata & \nodata & \nodata & \nodata & \nodata & \nodata \\
2M18093270-2046243 & $-$96.398 & 0.161 & \nodata & \nodata & \nodata & \nodata & \nodata & \nodata & \nodata & \nodata & \nodata \\
2M18093410-2047235 & $-$82.504 & 0.242 & \nodata & $-$0.966 & $-$0.837 & \nodata & \nodata & \nodata & \nodata & \nodata & \nodata\\
2M18093595-2045134 & $-$81.327 & 0.173 & \nodata & \nodata & \nodata & \nodata & \nodata & \nodata & \nodata & \nodata & \nodata\\
2M18093761-2047483 & $-$79.247 & 0.302 & \nodata & \nodata & \nodata & $-$0.874 & $-$0.172 & 1.083 & $-$0.752 & 0.213 & 0.619\\
2M18093935-2046378 & $-$87.985 & 0.153 & \nodata & \nodata & \nodata & $-$1.019 & $-$0.200 & 0.860 & 0.344 & 0.058 & 0.217\\
\textbf{Palomar 6} & & & & &  & & & & & \\
2M17433738-2612050 & 170.807 & 0.724 & 7.271 & $-$9.275 & $-$5.480 & $-$0.767 & $-$0.192 & 0.891 & 0.391 & 0.274 & 0.278\\
2M17433806-2613426 & 170.652 & 0.640 & 6.370 & $-$9.328 & $-$4.922 & $-$0.911 & $-$0.136 & 0.343 & 0.159 & 0.283 & 0.078\\
2M17434071-2613528 & 178.616 & 0.501 & 7.292 & $-$9.114 & $-$5.707 & $-$1.070 & 0.008 & 0.111 & \nodata & 0.344 & $-$0.147\\
2M17434331-2610217 & 175.957 & 1.177 & \nodata & $-$9.302 & $-$5.258 & $-$0.887 & $-$0.069 & 0.124 & 0.310 & 0.379 & 0.070\\
2M17434675-2616068 & 175.716 & 1.480 & 6.710 & $-$9.351 & $-$5.330 & $-$0.799 & $-$0.083 & 0.699 & 0.360 & 0.309 & 0.168\\
\textbf{Terzan 2} & & & & &  & & & & & \\ 
2M17273185-3048156 & 132.974 & 0.366 & \nodata & $-$2.094 & $-$6.309 & $-$0.763 & $-$0.085 & 0.909 & 0.047 & 0.259 & 0.282\\
2M17273364-3047243 & 131.310 & 0.835 & \nodata & $-$2.170 & $-$6.459 & $-$0.775 & $-$0.101 & 0.909 & 0.290 & 0.340 & 0.202\\
2M17273419-3048097 & 133.152 & 0.405 & \nodata & $-$2.079 & $-$6.194 & $-$0.874 & $-$0.057 & 0.619 & 0.044 & 0.329 & 0.042\\
2M17273540-3047308 & 135.249 & 0.967 & \nodata & $-$2.191 & $-$6.414 & $-$0.877 & $-$0.007 & 0.115 & $-$0.018 & 0.422 & 0.067\\
\textbf{Terzan 4} & & & & & & & & & & \\
2M17302949-3135089 & $-$47.944 & 1.917 & \nodata & $-$4.991 & $-$3.480 & $-$1.414 & $-$0.285 & 0.528 & 0.147 & 0.248 & $-$0.019\\
2M17303796-3135329 & $-$52.774 & 0.309 & \nodata & $-$5.159 & $-$3.719 & $-$1.292 & $-$0.255 & 0.867 & 0.519 & 0.011 & 0.721\\
2M17303838-3135492 & $-$44.255 & 0.117 & \nodata & $-$5.488 & $-$3.644 & $-$1.442 & $-$0.314 & 0.214 & \nodata & 0.250 & $-$0.258\\
\textbf{Liller1} & & & & & & & & & & \\
2M17332090-3322320 & 68.586 & 1.425 & 4.249 & $-$5.374 & $-$7.245 & \nodata & \nodata & \nodata & \nodata & \nodata & \nodata\\
2M17332272-3323206 & 71.754 & 0.494 & \nodata & $-$4.551 & $-$7.259 & \nodata & \nodata & \nodata & \nodata & \nodata & \nodata\\
2M17332472-3323166 & 56.550 & 0.119 & \nodata & $-$5.805 & $-$7.250 & \nodata & \nodata & \nodata & \nodata & \nodata & \nodata\\
2M17332881-3322499 & 49.152 & 1.284 & \nodata & $-$5.645 & $-$7.714 & $-$0.578 & 0.149 & 0.319 & \nodata & 0.305 & $-$0.036\\
\end{tabular}
\end{scriptsize}
\end{table*}

\begin{table*}
\begin{scriptsize}
\centering
\caption{Statistical properties of the APOGEE velocity distributions 
\label{stats}}
\begin{tabular}{p{0.75in}p{0.5in}p{0.5in}p{0.35in}p{0.35in}p{1.1in}p{0.8in}p{1.1in}p{1.1in}} \\ \hline
Field & $l$ range & $b$ range & Num Stars & \# Gaussians & mean & sigma & weight & data \\ \hline
($l$,$b$)=($-$10,0) & $-$11:$-$9 & $-$1:1 & 257 & 2 & $-$71, $-$177 & 59, 67 & 0.70, 0.30 & APOGEE DR16\\
($l$,$b$)=($-$5.5,0) & $-$6:$-$5 & $-$1:1 & 312 & 4 & $-$45, $-$133, 53, $-$257 & 36, 31, 43, 31 & 0.34, 0.31, 0.21, 0.14  & APOGEE + AAT \\ 
($l$,$b$)=($-$5.5,0) & $-$6:$-$5 & $-$1:1 & 220 & 4 & $-$31, $-$122, 57, $-$256 & 29, 34, 34, 30 & 0.29, 0.33, 0.22, 0.15  & APOGEE \\ 
($l$,$b$)=($-$5.8,0) & $-$6:$-$5.5 & $-$1:1  & 131 & 4 & $-$53, $-$140, 49, $-$260 & 37, 29, 54, 33 & 0.39, 0.32, 0.17, 0.12  & APOGEE + AAT \\ 
($l$,$b$)=($-$5.3,0) & $-$5.5:$-$5 & $-$1:1  & 181 & 4 & $-$121, $-$30, 57, $-$255 & 34, 29, 35, 30 & 0.34, 0.26, 0.24, 0.16  & APOGEE DR16 \\ 
($l$,$b$)=($-$5.5,$-$2) & $-$6:$-$5 & $-$3:$-$1 & 116 & 2 & $-$25, $-$101 & 73, 162 & 0.70, 0.30 & APOGEE DR16 \\ 
($l$,$b$)=($-$5.5,2) & $-$6:$-$5 & 1:3  & 86 & 3 & $-$87, 43, $-$254 & 43, 56, 39 &  0.55, 0.32, 0.13 & APOGEE DR16 \\
($l$,$b$)=($-$4.5,0) & $-$5:$-$4 & $-$1:1 & 245 & 2 & $-$24, $-$209 & 76, 76 & 0.72, 0.28 & APOGEE DR16\\
($l$,$b$)=($-$3.5,0) & $-$4:$-$3 & $-$1:1 & 80 & 3 & $-$62, $-$222, 94 & 54, 62, 46 & 0.42, 0.37, 0.21 & APOGEE DR16\\
($l$,$b$)=($-$2,0) & $-$3:$-$1 & $-$1:1 & 401 & 4 & $-$99, 4, 129, $-$228 & 47, 47, 74, 52 & 0.37, 0.33, 0.16, 0.15 & APOGEE DR16\\
($l$,$b$)=($+$2,0) & 1:3 & $-$1:1 & 477 & 3 & 8.7, 148, $-$126 & 53, 71, 60 & 0.49, 0.30, 0.21 & APOGEE DR16\\
($l$,$b$)=($+$3.5,0) & 3:4 & $-$1:1 & 67 & 2 & 2, 144 & 62, 77 & 0.73, 0.27 & APOGEE DR12/14\\
($l$,$b$)=($+$4.5,0) & 4:5 & $-$1:1 & 187 & 2 & 32, 108 & 83, 160 & 0.72, 0.28 & APOGEE DR12/14\\
($l$,$b$)=($+$5.3,0) & 5:5.5 & $-$1:1 & 126 & 3 & 18, 144, $-$137 & 44, 80, 119 & 0.66, 0.29, 0.05 & APOGEE DR12/14\\
($l$,$b$)=($+$5.8,0) & 5.5:6 & $-$1:1 & 126 & 2 & 19, 233 & 71, 50 & 0.79, 0.21 & APOGEE DR12/14\\
($l$,$b$)=($+$5.5,0) & 5:6 & $-$1:1 & 252 & 2 & 24, 236 & 75, 43 & 0.86, 0.14 & APOGEE DR12/14\\
($l$,$b$)=($+$5.5,$-$2) & 6:5 & $-$3:$-$1 & 899 & 2 & 16, 92 & 74, 112 & 0.63, 0.37 & APOGEE DR16 \\ 
($l$,$b$)=($+$5.5,2) & 6:5 & 1:3  & 618 & 2 & 8.2, 117 & 82, 94 & 0.63, 0.37 & APOGEE DR16 \\
($l$,$b$)=($+$10,0) & 9:11 & $-$1:1 & 253 & 2 & 50, 158 & 57, 125 & 0.75, 0.25 & APOGEE DR12/14\\
\end{tabular}
\end{scriptsize}
\end{table*}
{}

\end{document}